\RequirePackage{fix-cm} 
\documentclass[a4paper, twoside, reqno, dvips, 12pt]{amsart}
\usepackage{fixltx2e}   

\usepackage{etex}

\usepackage[latin1]{inputenc}
\usepackage[T1]{fontenc}

\usepackage{eucal}
\usepackage{esint}
\usepackage{dsfont}
\usepackage{xspace}
\usepackage{amsgen}
\usepackage{amsthm}
\usepackage{amssymb}
\usepackage{amsmath}
\usepackage{wasysym}
\usepackage{amsfonts}
\usepackage{mathtools}
\usepackage{MnSymbol}

\usepackage{sansmath}
\usepackage{mathrsfs}
\DeclareMathAlphabet{\mathscrbf}{OMS}{mdugm}{b}{n}

\usepackage{a4wide}

\usepackage[dvipsnames, table]{xcolor}
\definecolor{bckg}{RGB}{20.8, 20.8, 20.8}
\definecolor{oneblue}{rgb}{0.0, 0.0, 0.85}
\definecolor{Lightblue}{RGB}{214, 214, 214}
\definecolor{bluepigment}{rgb}{0.2, 0.2, 0.6}
\definecolor{charcoal}{rgb}{0.21, 0.27, 0.31}
\definecolor{denimblue}{rgb}{0.08, 0.38, 0.74}
\definecolor{Lightgray}{rgb}{0.89, 0.89, 0.89}
\definecolor{darkgrey}{rgb}{0.273, 0.281, 0.30}
\definecolor{darkelectricblue}{rgb}{0.33, 0.41, 0.47}

\usepackage{psfrag}
\usepackage{graphicx}
\usepackage{subfigure}
\usepackage{morefloats}
\usepackage{indentfirst}

\usepackage{acronym}
\usepackage{microtype}
\usepackage[labelsep=period,%
            labelfont={bf,sf,color=bluepigment},%
            justification=raggedright]{caption}

\usepackage[perpage, symbol]{footmisc}

\usepackage[usenames, dvipsnames]{pstricks}
\usepackage{epsfig}
\usepackage{pst-grad} 
\usepackage{pst-plot} 

\usepackage[colorlinks,
           urlcolor=oneblue,
           linkcolor=denimblue,
           citecolor=NavyBlue,
           bookmarksopen=false,
           pdfpagemode=UseNone,
           pagebackref]{hyperref}

\usepackage[sort&compress, comma, square, numbers]{natbib}

\usepackage[explicit]{titlesec}

\titleformat{\section}
  {\color{NavyBlue}\Large\sffamily\bfseries}
  {}
  {0em}
  {\colorbox{bckg!5}{\parbox{\dimexpr\linewidth-2\fboxsep\relax}{\centering\thesection. #1}}}
  [\vspace*{0.33em}]

\titleformat{name=\section,numberless}
  {\color{NavyBlue}\Large\sffamily\bfseries}
  {}
  {0.0em}
  {\colorbox{bckg!10}{\parbox{\dimexpr\linewidth-2\fboxsep\relax}{\centering#1}}}
  [\vspace*{0.33em}]

\titleformat{\subsection}
  {\color{NavyBlue}\large\sffamily\bfseries}
  {}
  {0.0em}
  {\colorbox{bckg!5}{\parbox{\dimexpr\linewidth-2\fboxsep\relax}{\centering\thesubsection. #1}}}
  [\vspace*{0.33em}]

\titleformat{name=\subsection,numberless}
  {\color{NavyBlue}\Large\sffamily\bfseries}
  {}
  {0em}
  {\colorbox{bckg!10}{\parbox{\dimexpr\linewidth-2\fboxsep\relax}{\centering#1}}}
  [\vspace*{0.33em}]

\titleformat{\subsubsection}
  {\color{bluepigment}\sffamily\normalsize\bfseries}
  {\thesubsubsection}
  {0.5em}
  {#1}
  [\vspace*{0.33em}]

\titleformat{\paragraph}[runin]
  {\color{bluepigment}\sffamily\small\bfseries}
  {}
  {0em}
  {#1}

\titlespacing{\section}{1.0em}{1.5em plus 2pt minus 2pt}%
{1.0em plus 2pt minus 2pt}[0em]
\titlespacing{\subsection}{1.0em}{1.5em plus 2pt minus 2pt}%
{1.0em}[0em]
\titlespacing{\subsubsection}{1.0em}{1.5em plus 2pt minus 2pt}%
{1.0em plus 2pt minus 2pt}[0em]

\usepackage{titletoc}

\contentsmargin{0.5em}
\newlength{\tocsep} 
\titlecontents{section}[\tocsep]
  {\addvspace{10pt}\bfseries\sffamily}
  {\contentslabel[\thecontentslabel]{\tocsep}}
  {}
  {\ \titlerule*[0.75pc]{.}\ \thecontentspage}
  []
\titlecontents{subsection}[\tocsep]
  {\addvspace{8pt}\sffamily}
  {\contentslabel[\thecontentslabel]{\tocsep}}
  {}
  {\ \titlerule*[0.5pc]{.}\ \thecontentspage}
  []
\titlecontents*{subsubsection}[\tocsep]
  {\addvspace{2pt}\footnotesize\sffamily}
  {}
  {}
  {\ \titlerule*[0.35pc]{.}\ \thecontentspage}
  [\\*]

\makeatletter
\def\@setauthors{%
  \begingroup
  \def\thanks{\protect\thanks@warning}%
  \trivlist
  \centering\footnotesize \@topsep30\p@\relax
  \advance\@topsep by -\baselineskip
  \item\relax
  \author@andify\authors
  \def\\{\protect\linebreak}%
  \textsc{\normalsize\textcolor{darkelectricblue}{\authors}}%
  \ifx\@empty\contribs
  \else
    ,\penalty-3 \space \@setcontribs
    \@closetoccontribs
  \fi
  \endtrivlist
  \endgroup
}
\def\@settitle{\begin{center}%
  \baselineskip14\p@\relax
    \bfseries
    \textsc{\Large\textcolor{charcoal}{\@title}}
  \end{center}%
}
\makeatother

\usepackage{enumitem}
\setlist[description]{%
  topsep=30pt,               
  itemsep=5pt,               
  font={\bfseries\sffamily\color{NavyBlue}}, 
}

\usepackage{fancyhdr}
\usepackage{lastpage}

\newcommand*\Title{\textcolor{bluepigment}{Computation of steady surface gravity waves}}
\newcommand*\Authors{\textcolor{bluepigment}{D.~Clamond \& D.~Dutykh}}
\newcommand*{\plogo}{\textcolor{gray}{{\texttt{arXiv.org} / \textsc{hal}}}} 

\numberwithin{equation}{section}

\newtheorem{remark}{Remark}

\newcommand{\up}[1]{$^{\mathrm{\small\textsf{#1}}}$} 

\newcommand{\depth}{d}                               
\newcommand{\upi}{{\pi}}
\newcommand{\ud}{\mathrm{d}}

\newcommand{\ue}{\mathrm{e}}
\newcommand{\ui}{\mathrm{i}}
\renewcommand{\O}{\mathcal{O}}

\newcommand{\sur}[1]{\tilde{#1}}                     
\renewcommand{\bot}[1]{\bar{#1}}                     
\newcommand{\Sur}[1]{\widetilde{#1}}                 
\newcommand{\Bot}[1]{\overline{#1}}                  

\newcommand{\cs}{c_{\,\textsc{s}}}
\newcommand{\ce}{c_{\,\textsc{e}}}
\newcommand{\cl}{c_{\,\textsc{r}}}

\newcommand{\kbar}{\mathchar'26\mkern-9mu k}
\newcommand{\dbar}{\;\mathchar'26\mkern-14mu d\/}

\renewcommand{\Re}{\operatorname{Re}}
\renewcommand{\Im}{\operatorname{Im}}

\newcommand{\eqdef}{\mathop{\stackrel{\,\mathrm{def}}{:=}\,}}

\newcommand{\cf}{\emph{cf.}\xspace}
\newcommand{\ie}{\emph{i.e.}\xspace}
\newcommand{\eg}{\emph{e.g.}\xspace}
\newcommand{\etc}{\emph{etc.}\xspace}

\newcommand{\half}{{\textstyle{1\over2}}}
\newcommand{\halfi}{{\textstyle{1\over2\ui}}}
\newcommand{\threehalf}{{\textstyle{3\over2}}}

\begin{document}

\title[\Title]{Accurate fast computation of steady two-dimensional surface gravity waves in arbitrary depth}

\author[D.~Clamond]{Didier Clamond$^*$}
\address{\textbf{D.~Clamond:} Universit\'e C\^ote d'Azur, CNRS--LJAD UMR 7351,
Parc Valrose, F-06108 Nice, France}
\email{didierc@unice.fr}
\urladdr{http://math.unice.fr/~didierc/}
\thanks{$^*$ Corresponding author}

\author[D.~Dutykh]{Denys Dutykh}
\address{\textbf{D.~Dutykh:}  Univ. Grenoble Alpes, Univ. Savoie Mont Blanc, CNRS, LAMA, 73000 Chamb\'ery, France and LAMA, UMR 5127 CNRS, Universit\'e Savoie Mont Blanc, Campus Scientifique, F-73376 Le Bourget-du-Lac Cedex, France}
\email{Denys.Dutykh@univ-savoie.fr}
\urladdr{http://www.denys-dutykh.com/}


\begin{titlepage}
\thispagestyle{empty} 
\noindent
{\Large Didier \textsc{Clamond}}\\
{\it\textcolor{gray}{Universit\'e C\^ote d'Azur, LJAD, France}}
\\[0.02\textheight]
{\Large Denys \textsc{Dutykh}}\\
{\it\textcolor{gray}{CNRS, Universit\'e Savoie Mont Blanc, France}}
\\[0.16\textheight]

\colorbox{Lightblue}{
  \parbox[t]{1.0\textwidth}{
    \centering\huge\sc
    \vspace*{1.09cm}
    
    \textcolor{bluepigment}{Accurate fast computation of steady two-dimensional surface gravity waves in arbitrary depth}
    
    \vspace*{0.5cm}
  }
}

\vfill 

\raggedleft     
{\large \plogo} 
\end{titlepage}


\newpage
\thispagestyle{empty} 
\par\vspace*{\fill}   
\begin{flushright} 
{\textcolor{denimblue}{\textsc{Last modified:}} \today}
\end{flushright}


\newpage
\maketitle
\thispagestyle{empty}


\begin{abstract}

This paper describes an efficient algorithm for computing steady two-di\-men\-sional surface gravity wave in irrotational motion. The algorithm complexity is $\O\,(N\log N)\,$, $N$ being the number of \textsc{Fourier} modes. This feature allows the arbitrary precision computation of waves in arbitrary depth, \ie, it works efficiently for \textsc{Stokes}, cnoidal and solitary waves, even for quite large steepnesses, up to about ninety-nine percent of the maximum steepness for all wavelengths. In particular, the possibility to compute accurately very long (cnoidal) waves is a feature not shared by other algorithms and asymptotic expansions. The method is based on conformal mapping, \textsc{Babenko} equation rewritten in a suitable way, pseudo-spectral method and \textsc{Petviashvili}'s iterations. The efficiency of the algorithm is illustrated via some relevant numerical examples. The code is open source, so interested readers can easily check the claims, use and modify the algorithm.

\bigskip
\noindent \textbf{\keywordsname:} Periodic waves; cnoidal waves; Stokes wave; Babenko equation; Petviashvili scheme; spectral methods \\


\end{abstract}


\newpage
\tableofcontents
\thispagestyle{empty}


\newpage
\section{Introduction}

Many physical phenomena and mathematical problems related to surface gravity water waves remain unknown, not well-understood or unsolved, even in the `simple' case of traveling waves of permanent form in two-dimensional irrotational motion \cite{Clamond2012a, Constantin2012}. Traveling waves are of special interest because complex sea states are often described as superposition and interaction of such waves. Surveys of analytical and numerical models, their limitations and open questions can be found in dedicated articles \cite{Dias1999, Fenton1988, Fenton1999a, Groves2004, Strauss2010} and books \cite{Constantin2011a, Okamoto2001, Vanden-Broeck2010}.

Since exact analytic solutions for irrotational steady surface gravity waves are still unknown, and likely will never be known, only analytic or numerical approximations are accessible. Simple analytic approximations are interesting for physical insights, but they are of limited accuracy. Even formal analytic solutions in terms of small parameter expansions have limited accuracy because they generally converge slowly \citep{Schwartz1974}, when they converge (shallow water expansions are divergent for all amplitudes \citep{Germain1967a}). 
Thus, their numerical calculation suffer large truncation errors, and are prone to important accumulation of round-off and cancelation errors. Even when a simple analytic approximation would be sufficient for a given application, its computation may be practically intractable. An example is KdV cnoidal wave analytic solution that cannot be easily computed for very long waves, as demonstrated in the present paper. Therefore, only numerical approximations of the original equations can provide highly accurate solutions that are necessary for many applications. For instance, for stability analysis and interactions of traveling waves using accurate numerical models, a too crude (about six digits accuracy, say) initial condition for traveling wave may lead to incorrect behaviours, specially for long-time simulations. A lack of initial accuracy may then lead to erroneous physical interpretations. Another example is the numerical investigation of mathematical conjectures. With arbitrary precision computations, one can check if a conjecture is likely true or not, or can formulate new conjectures worthy investigations. Indeed, some questions of mathematical interest (\eg, rigorous proofs of unicity and monotonicity regarding velocity, acceleration and pressure fields) remain open for all waves and not only for very large amplitudes (see the papers in \cite{Constantin2012} for reviews).

Several algorithms have been proposed in the literature for the computation of steady surface waves solutions of the irrotational \textsc{Euler} equations \citep{Okamoto2001}. The focus was more on the computation of the almost highest gravity waves (see, \eg, \cite{Byatt-Smith2001, Maklakov2002, Williams1985}) or `exotic' capillary-gravity waves (\eg, \cite{Clamond2015a, Vanden-Broeck2010}), than the computation of arbitrary wavelengths. Actually, none of these algorithms are capable of computing long waves in shallow water, not even the ones of small amplitude, because they are too demanding. Indeed, all these algorithms lead to the resolution of a discrete system of nonlinear equations, this system being large for long waves. The resolution of this system is generally performed with (the like of) \textsc{Newton} or, better, \textsc{Levenberg--Marquardt} iterations \cite{Lourakis2005}. Though robust and effective, these methods are computationally very demanding because each iteration requires $\O\,(N^{\,3})$ operations, $N$ being the number of unknowns (\eg the \textsc{Fourier} coefficient for periodic water waves). Indeed, though their theoretical minimum complexity is $\O\,(N^{\,2})\,$, usually \textsc{Levenberg--Marquardt} iterations use direct solvers that have $\O(N^{\,3})$ complexity. In the simplest case, they rely on \textsc{Cholesky} decomposition which needs $\O\,(N^{\,3})$ operations. In most numerically robust version of the \textsc{Levenberg}--\textsc{Marquardt} method, its implementations relies on rank-revealing QR algorithm, which is also $\O\,(N^{\,3})$ (with a big constant in front of $N^{\,3}$). Thus, when $N$ is very large, the computational time may be prohibitive and the accumulation of round-off errors significant, even if an algorithm of complexity $\O\,(N^{\,2})$ can be used. It is well-known that for steep waves and cnoidal waves in shallow water, the number of \textsc{Fourier} modes needed for accurate resolutions must be very large, specially when using conformal mapping. Therefore, an algorithm with complexity lower than $\O\,(N^{\,2})$ is desirable to achieve these computations. It is the purpose of the present paper to describe such an algorithm with overall complexity $\O\,(N\log N)\,$, thus suitable for arbitrary precision computations of all waves of practical interest in arbitrary depth.

For a simple wave equation, \cite{Petviashvili1976} proposed an algorithm based on stabilised fixed point iterations for computing solitary waves. This algorithm is interesting because it is very simple to implement and has complexity $\O\,(N)\,$. However, \textsc{Petviashvili}'s method generally works only for equations with special nonlinear terms, typically autonomous equations with homogeneous nonlinearities \citep{Yang2010}. Several variants have then been proposed to extend somehow the scope of these modified \textsc{Petviashvili} methods \cite{Ablowitz2005, Alvarez2014a, Lakoba2007}. Instead of tweaking the algorithm, our approach here is to rewrite the \textsc{Euler} equations in a form suitable for their numerical resolution via the classical \textsc{Petviashvili} method. Doing so, we obtain an algorithm suitable to compute waves in arbitrary depth, in particular long waves with wavelength of tenths to millions times the mean water depth. To our knowledge, it is the first algorithm capable of computing such long waves.

The \textsc{Petviashvili} method works for the \textsc{Euler} equations if they are rewritten in the form of a \textsc{Babenko} equation \citep{Babenko1987}. This was successfully implemented to compute solitary gravity waves \cite{Clamond2012b, Dutykh2013b}. Unfortunately, this algorithm does not work for periodic waves. In infinite depth, periodic waves were successfully computed with a modified \textsc{Petviashvili} method by \cite{Dyachenko2014}. Their algorithm does not work in finite depth, however. In order to overcome these drawbacks, we propose here a simple change of variable that transforms the \textsc{Babenko} equation into a form tractable with the classical \textsc{Petviashvili} method. Our algorithm is not designed for the most extreme waves: for a given wavelength, it works only for all wave-heights less than about ninety-nine percent of the maximum one. Thus, we are able to rapidly compute waves of practical interest in arbitrary depth (infinite, finite and shallow) and to arbitrary precision.

The paper is organised as follow. In Section~\ref{secmathdef}, we present the physical assumptions and the mathematical definitions and notations. In Section~\ref{secconmap}, we introduce a conformal mapping and we give precise definitions of all the variables and parameters in the conformal space. In Section~\ref{secbab}, we derive a \textsc{Babenko} equation written in a suitable form for a fast numerical resolution. The numerical algorithm is subsequently described in Section~\ref{secnum} and relevant numerical examples are provided in Section~\ref{secexa}. Summary and perspectives are outlined in Section~\ref{secconc}.


\section{Definitions and notations}
\label{secmathdef}

We consider steady two-dimensional potential flows due to surface gravity waves in constant depth $\depth\,$. The fluid is of (positive) constant density $\rho\,$, the pressure is zero at the impermeable free surface, while the seabed is fixed, horizontal and impermeable.

\begin{figure}
  \centering
  \includegraphics[width=0.99\textwidth]{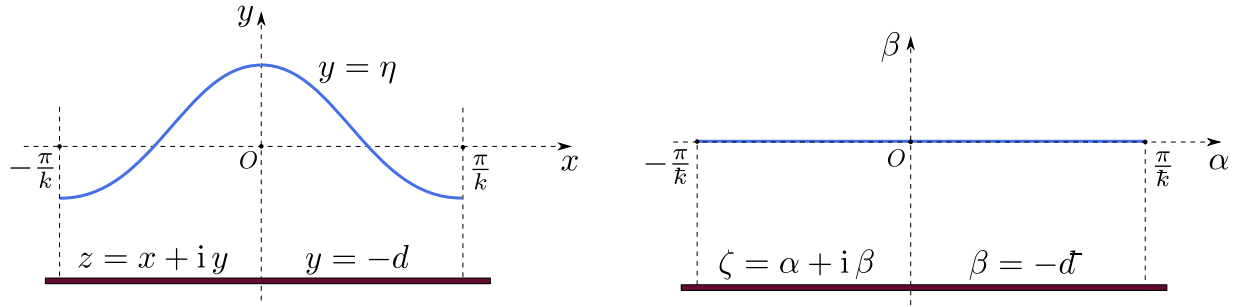}
  \caption{\small\em Definition sketch. Left: $z-$plane; right: $\zeta-$plane.}
  \label{figsketch}
\end{figure}

Let be $(x,\,y)$ a \textsc{Cartesian} coordinate system moving with the wave, $x$ being the horizontal coordinate and $y$ being the upward vertical one. The wave is ($2\upi/k$)-periodic\footnote{The fundamental wavenumber $k$ is zero for solitary and aperiodic waves.} and $x\ =\ 0$ is the abscissa of a wave crest. By $y\ =-\depth\,$, $y\ =\ \eta\,(x)$ and $y\ =\ 0$ we denote, respectively, the equations of the bottom, of the free surface and of the mean water level (see Figure~\ref{figsketch}, left). The latter implies that $\left<\eta\right>\ =\ 0$ --- $\left<\bullet\right>$ the \textsc{Eulerian} average operator over one spatial period (wavelength) --- \ie
\begin{equation}\label{defmean}
  \left<\,\eta\,\right>\ \eqdef\ {k\over2\,\upi}\int_{-{\pi/k}}^{\,{\pi/k}}\eta\,(x)\ \ud\/x\ =\ 0\,.
\end{equation}
$a\ \eqdef\ \eta\,(0)$ denotes the wave crest amplitude and $b\ \eqdef\ -\eta\,(\upi/k)$ is the wave trough amplitude, so that $H\ \eqdef\ a\ +\ b$ is the total wave height. A wave steepness $\varepsilon$ is then classically defined as $\varepsilon\ \eqdef\ k\/H/2\,$. Variable $\phi$ denoting the velocity potential, the classical equations of motion are
\begin{align*}
\phi_{\,x\,x}\ +\ \phi_{\,y\,y}\ &=\ 0 \qquad\mathrm{for}\quad -d\,\leqslant\, y\,\leqslant\,\eta\,(x)\,, \\
\phi_{\,y}\ &=\ 0 \qquad\mathrm{\,at}\qquad y\,=\,-d\,, \\
\phi_{\,y}\ -\ \eta_{\,x}\,\phi_{\,x}\ &=\ 0 \qquad\mathrm{\,at}\qquad y\,=\,\eta\,(x)\,, \\
2\,g\,\eta\ +\ \phi_{\,x}^{\,2}\ +\ \phi_{\,y}^{\,2}\ &=\ B \qquad\mathrm{\!at}\qquad y\,=\,\eta\,(x)\,,
\end{align*}
where $g\ >\ 0$ is the (constant) acceleration due to gravity and $B$ is a \textsc{Bernoulli} constant.

Let be $\psi\,$, $u$ and $v$ the stream function, the horizontal and vertical velocities, respectively, such that $u\ =\ \phi_{\,x}\ =\ \psi_{\,y}$ and $v\ =\ \phi_{\,y}\ =\ -\psi_{\,x}\,$. It is convenient to introduce the complex potential $f\ \eqdef\ \phi\ +\ \ui\,\psi$ (with $\ui^2\ =\ -1$) and the complex velocity $w\ \eqdef\ u\ -\ \ui\,v$ that are holomorphic functions of $z\ \eqdef\ x\ +\ \ui\,y$ (\ie, $f\ =\ f\,(z)$ and $w\ =\ \ud f/\ud z$). The complex conjugate is denoted with a star (\eg, $z^*\ =\ x\ -\ \ui\,y$), while over `bars' denote the quantities written at the seabed --- \eg, $\bot{z}\,(x)\ =\ x\ -\ \ui\,\depth\,$, 
$\bot{\phi}\,(x)\ =\ \phi\,(x,\,y\!=\!-\depth)$ --- and over `tildes' denote the quantities written at the surface --- \eg, $\sur{z}\,(x)\ =\ x\ +\ \ui\,\eta\,(x)\,$, $\sur{\phi}\,(x)\ =\ \phi\,(x,\,y\!=\!\eta\,(x))\,$.\footnote{Note that, \eg, $\sur{u}\ =\ \widetilde{\ \phi_{\,x}\,}\ \neq\ \sur{\phi}_{\,x}\ =\ \sur{u}\ +\ \eta_{\,x}\,\sur{v}\,$.} Free surface and bottom being streamlines, $\sur{\psi}$ and $\bot{\psi}$ are constants. One can then take $\sur{\psi}\ =\ 0$ \emph{or} $\bot{\psi}\ =\ 0$ without loss of generality (gauge condition for the stream function).

The pressure field can be obtained from the \textsc{Bernoulli} equation
\begin{equation}\label{bernbase}
  2\,p\ +\ 2\,g\,y\ +\ u^{\,2}\ +\ v^{\,2}\ =\ B\,,
\end{equation}
where $p$ is the pressure divided by the density. At the free surface the pressure being zero (\ie, $\sur{p}\ =\ 0$), the \textsc{Bernoulli} constant $B$ is defined averaging \eqref{bernbase} applied at the free surface and using the condition \eqref{defmean}, \ie
\begin{equation}\label{defB}
  B\ =\, \left<\,\sur{u}^{\,2}\,+\,\sur{v}^{\,2}\,\right>\,.
\end{equation}
From the incompressibility and the irrotationality, it follows that $B$ can also 
be obtained from the expression at the bottom \cite{Clamond2013b} 
\begin{equation}\label{defBot}
  B\ =\, \left<\,\bot{u}^2\,\right>\,,
\end{equation}
and then, from the \textsc{Bernoulli} equation averaged at the bed, gives $\left<\bot{p}\right>\ =\ g\depth\,$. More generally, $B$ equals $u^{\,2}\ +\ v^{\,2}$ averaged along any streamline (in the frame of reference moving with the wave).

Let be $-\cs$ the mean flow velocity defined as
\begin{equation}\label{defcs}
  \cs\ \eqdef\ -\left<\,\frac{1}{\depth}\,\int_{-\depth}^\eta\,u\,(x,\,y)\ \ud\/y\,\right>\ =\ \frac{\bot{\psi}\,-\,\sur{\psi}}{\depth}\,.
\end{equation}
Thus, $\cs$ is the phase velocity of the wave observed in the frame of reference without mean flow (\textsc{Stokes}' second definition of phase celerity), that is also the frame where the wave impulse is zero (Appendix~\ref{appint}). Another important quantity is the phase velocity $\ce$ observed in the frame of reference without mean velocity at the seabed (that is also the one where the circulation is zero, \cf~Appendix~\ref{appint}): 
\begin{equation}\label{defce}
  \ce\ \eqdef\  -\left<\, \bot{u}\,\right>\ =\ -\left<\, u\,(x,\,y\!=\!-d)\,\right>\,.
\end{equation}
This is \textsc{Stokes}' first definition of phase celerity. Since the motion is irrotational, $\ce$ can be obtained averaging $u$ along any horizontal line $y\ =\ \mathrm{constant}$ (but not along the wavy surface, \ie, $\ce\ \neq\ -<\sur{u}>$). Many other phase velocities can of course be defined, but $\cs$ and $\ce$ are two velocities of special interest here. Notice that $B\ =\ \cs^{\,2}\ =\ \ce^{\,2}$ in deep water and for solitary waves (see below), and that neither $\cs$ and $\ce$ are the linear phase velocity $c_{\,0}\ \eqdef\ \sqrt{(g/k)\tanh(k\depth)}$ if the wave amplitude is not zero. A discussion on the \textsc{Bernoulli} constant and phase velocities can be found in \cite{Clamond2017c}. Further informations regarding these phase velocities and integral quantities are given here in the Appendices~\ref{apppsi}, \ref{appint} and \ref{appaverel}.


\section{Conformal mapping}
\label{secconmap}

Let be the change of independent complex variable $z\ \mapsto\ \zeta\ \eqdef\ (\ui\,\sur{\psi}\ -\ f)/\cl\,$, $\cl\ \neq\ 0$ being a velocity of reference. In practice, one could take $\cl\ =\ \cs$ or $\cl\ =\ \ce$ or $\cl\ =\ c_{\,0}$ or $\cl\ =\ B^{\,1/2}$ or $\cl\ =\ (\ce\cs)^{\,1/2}\,$, for example, but another convenient choice can be made depending on the problem at hands. One should take $\cl\ >\ 0$ if the wave travels toward the increasing $x-$direction in a `fixed' frame of reference, and $\cl\ <\ 0$ is the wave travels toward the decreasing $x-$direction. Here, without loss of generality, we consider only waves travelling toward the increasing $x-$direction.

This change of variable conformally maps the fluid fundamental domain 
\begin{equation*}
  0\,\leqslant\,x\,\leqslant\,2\,\pi\,/\,k, \qquad -\,\depth\,\leqslant\,y\,\leqslant\,\eta\,(x)\,,
\end{equation*}
into the rectangle (see Figure~\ref{figsketch}, right)
\begin{equation*}
  0\,\leqslant\,\alpha\,\leqslant\,2\,\pi\,\ce\,/\,k\,\cl, \qquad -\,\depth\,\cs\,/\,\cl\,\leqslant\,\beta\,\leqslant\,0\,,
\end{equation*}
where $\alpha\ \eqdef\ \Re\,(\zeta)$ and $\beta\ \eqdef\ \Im\,(\zeta)\,$. For convenience, we introduce the apparent wavenumber $\kbar$ and apparent depth $\dbar$ in the conformal plane   
\begin{equation*}
  \kbar\ \eqdef\ \cl\,\ce^{\,-1}\,k\,, \qquad
  \dbar\ \eqdef\ \cs\,\cl^{\,-1}\,d\,, 
\end{equation*}
that are generally different from the corresponding quantities in the physical plane. Note that $\cl$ does not appear in the expression of $\kbar\dbar\ =\ \cs\/\ce^{\,-1}\/k\/d\,$, so no choice of $\cl$ can enforce the equality $\kbar\dbar\ =\ kd\,$, the latter being obtained only if $\ce\ =\ \cs\,$. Conversely, the choice $\cl\ =\ \sqrt{\ce\cs}$ yields $\dbar/\kbar\ =\ \depth/k$ so, with this peculiar choice of $\cl\,$, the areas of the fundamental periods are identical in physical and conformal planes. For our numerical resolution, we found convenient to take 
$\cl\ =\ \ce$ (see Section~\ref{secnum} below).

Since $\ud z/\ud\zeta\ =\ z_{\,\alpha}\ =\ -\ui\,z_{\,\beta}$ (subscripts denoting partial derivatives), we have the \textsc{Cauchy--Riemann} relations $x_{\,\alpha}\ =\ y_{\,\beta}$ and $x_{\,\beta}\ =\ -y_{\,\alpha}\,$, while the complex velocity and the velocity components are
\begin{multline}\label{wuvqmap}
  \frac{w}{\cl}\, =\, -\left(\frac{\ud\,z}{\ud\/\zeta}\right)^{\!-1}\,, \qquad
  \frac{u}{\cl}\, =\, \frac{-\,x_{\,\alpha}}{x_{\,\alpha}^{\,2}\ +\ y_{\,\alpha}^{\,2}}\,, \\
  \frac{v}{\cl}\, =\, \frac{-\,y_{\,\alpha}}{x_{\,\alpha}^{\,2}\ +\ y_{\,\alpha}^{\,2}}\,, \qquad
  \frac{u^{\,2}\ +\ v^{\,2}}{\cl^{\,2}}\ =\ \frac{1}{x_{\,\alpha}^{\,2}\ +\ y_{\,\alpha}^{\,2}}\,
\end{multline}
thence
\begin{equation*}
  x_{\,\alpha}\ =\ y_{\,\beta}\ =\ \frac{-\,\cl\,u}{u^{\,2}\ +\ v^{\,2}}\,, \qquad
  y_{\,\alpha}\ =\ -x_{\,\beta}\ =\ \frac{-\,\cl\,v}{u^{\,2}\ +\ v^{\,2}}\,. 
\end{equation*}
With these relations one can compute all the physical quantities of interest.


\subsection{Conformal averaging operator}

The \textsc{Bernoulli} constant, from the relation \eqref{defB}, is defined in the conformal plane by
\begin{equation} \label{bernBxy}
  \frac{B}{\cl^{\,2}}\ =\ \frac{k}{2\,\upi}\;\int_{-\frac{\upi\ce}{k\cl}}^{\frac{\upi\ce}{k\cl}}\,\frac{x_{\,\alpha}}{x_{\,\alpha}^{\,2}\ +\ y_{\,\alpha}^{\,2}}\ \ud\/\alpha\ =\ \frac{\ce}{\cl}\;\frac{\kbar}{2\,\upi}\;\int_{-\upi/\kbar}^{\upi/\kbar}\,\frac{x_{\,\alpha}}{x_{\,\alpha}^{\,2}\ +\ y_{\,\alpha}^{\,2}}\ \ud\/\alpha\,, 
\end{equation}
the integral being computed keeping $\beta$ constant, in particular at the free surface $\beta\ =\ 0\,$. The relation \eqref{bernBxy} shows that it is convenient to introduce the average operator over one period in the $\alpha-$variable ($\beta$ being kept constant)
\begin{equation}\label{defmeanalpha}
  \left\llangle(\cdots)\right\rrangle\ \eqdef\ \frac{\kbar}{2\,\upi}\;\int_{-\upi/\kbar}^{\upi/\kbar}\,(\cdots)\ \ud\/\alpha\,,
\end{equation}
for any quantity $(\cdots)\,$. Thus, at the free surface and at the bottom we have respectively
\begin{align}
  \left\llangle\Sur{(\cdots)}\right\rrangle\ &=\ \frac{\kbar}{2\,\upi}\;\int_{-\upi/\kbar}^{\upi/\kbar}\,(\cdots)_{\,\beta\,=\,0}\ \ud\/\alpha\ =\ -\,\ce^{\,-1}\left<\,(\cdots)_{\,y\,=\,\eta}\,(\sur{u}+\sur{v}\/ \eta_x)\,\right>\ =\ -\,\ce^{\,-1}\left<\,\Sur{(\cdots)}\,\sur{\phi}_{\,x}\,\right>,\label{defmeanalphasur}\\
  \left\llangle\Bot{(\cdots)}\right\rrangle\ &=\ \frac{\kbar}{2\,\upi}\;\int_{-\upi/\kbar}^{\upi/\kbar}\,(\cdots)_{\,\beta\,=\,-\dbar}\ \ud\/\alpha\ =\ -\,\ce^{\,-1}\left<\,(\cdots)_{\,y\,=\,-\depth}\,\bot{u}\,\right>\ =\ -\,\ce^{\,-1}\left<\,\Bot{(\cdots)}\,\bot{\phi}_{\,x}\,\right>\,, \label{defmeanalphabot}
\end{align}
and conversely
\begin{align}
  \left<\Sur{(\cdots)}\right>\ &=\ \frac{k}{2\,\upi}\;\int_{-\upi/k}^{\upi/k}\,(\cdots)_{\,y\,=\,\eta}\ \ud\/x\ =\ -\,\ce\left\llangle\,(\cdots)_{\,\beta\,=\,0}\,\sur{u}\left/\left(\sur{u}^2+\sur{v}^2\right)\right.\right\rrangle,\label{defmeanxsur}\\
  \left<\Bot{(\cdots)}\right>\,&=\ \frac{k}{2\,\upi}\;\int_{-\upi/k}^{\upi/k}\,(\cdots)_{\,y\,=\,-\depth}\ \ud\/x\ =\ -\,\ce\left\llangle\,(\cdots)_{\,\beta\,=\,-\dbar}\,/\,\bot{u}\,\right\rrangle\,. \label{defmeanxbot}
\end{align}
Averaged physical quantities being defined in the physical plane and not in the conformal plane, the connections between the averaging operators are useful to express these physical quantities in the conformal plane. Conversely, these connections are also useful to express averaged quantities in the conformal plane (needed for a numerical resolution) in their physical plane counterparts. For an easy reference, we give several such relations in the Appendix~\ref{appaverel}.


\subsection{Resolution of the conformal mapping}

With the change of dependent variables
\begin{equation}
  x\ =\ \cl\,\ce^{\,-1}\,\alpha\ +\ X\,(\alpha,\,\beta)\,, \qquad
  y\ =\ \cl\,\ce^{\,-1}\,(\/\beta\ +\ \dbar\/)\ -\ \depth\ +\ Y\,(\alpha,\,\beta)\,, \label{defxyXY}
\end{equation}
the \textsc{Cauchy}--\textsc{Riemann} relations $X_{\,\alpha}\ =\ Y_{\,\beta}$ and $X_{\,\beta}\ =\ -Y_{\,\alpha}$ hold, while the bottom ($\beta\ =\ -\dbar$) and the free surface ($\beta\ =\ 0$) impermeabilities yield
\begin{equation*}
  \bot{Y}\,(\alpha)\ \eqdef\ Y\,(\alpha,\,-\dbar)\ =\ 0\,, \qquad 
  \sur{Y}\,(\alpha)\ \eqdef\ Y\,(\alpha,\,0)\ =\ \sur{y}\ +\ \depth\,(1\ -\ \cs/\ce)\,.
\end{equation*}
At the boundaries of the fundamental period (\ie, $\alpha\ =\ 0$ and $\alpha\ =\ 2\,\pi/\kbar$), we have from \eqref{defxyXY}({\it a})
\begin{equation*}
  X\,(0,\,\beta)\ =\ 0\,, \qquad 
  X\,(2\pi/\kbar,\,\beta)\ =\ 0\,,
\end{equation*}
and more generally $X\,(\alpha\ +\ 2\pi/\kbar,\,\beta)\ =\ X\,(\alpha,\,\beta)\,$. Therefore, the function $X$ is $(2\pi/\kbar)-$periodic.

The functions $X$ and $Y$ can be expressed in term of $\bot{X}$ --- \ie, the function $X$ written at the bottom --- as (see \cite{Clamond1999, Clamond2003} for detailed derivations)
\begin{align}
  X\,(\alpha,\,\beta)\ =&\ \half\,\bot{X}\,(\zeta\ +\ \ui\/\dbar)\ +\ \half\,\bot{X}\,(\zeta^*\ -\ \ui\/\dbar)\ =\ \cos\,\left[\/(\beta\ +\ \dbar)\,\partial_{\,\alpha}\/\right]\,\bot{X}\,(\alpha)\nonumber\\
  =&\ \sum_{n\,=\,0}^\infty\frac{(-1)^{\,n}\,(\beta\ +\ \dbar)^{\,2\,n}}{(2\,n)!}\,\frac{\partial^{\,2\,n}\,\bot{X}\,(\alpha)}{\partial\alpha^{\,2\,n}}, \label{solxxbot} \\
  Y\,(\alpha,\,\beta)\ =&\ \halfi\,\bot{X}\,(\zeta\ +\ \ui\/\dbar)\ -\ \halfi\,\bot{X}\,(\zeta^*\ -\ \ui\/\dbar)\ =\ \sin\,\left[\/(\beta\ +\ \dbar)\,\partial_{\,\alpha}\/\right]\,\bot{X}\,(\alpha)\nonumber\\
  =&\ \sum_{n\,=\,1}^\infty\frac{(-1)^{\,n\,+\,1}\,(\beta\ +\ \dbar)^{\,2\,n\,-\,1}}{(2\,n\,-\,1)!}\,\frac{\partial^{\,2\,n\,-\,1}\,\bot{X}\,(\alpha)}{\partial\alpha^{\,2\,n\,-\,1}}\,, \label{solyxbot}
\end{align}
where a star denotes the complex conjugate. Thus, the \textsc{Cauchy--Riemann} relations and the bottom impermeability are fulfilled identically. At the free surface $\beta\ =\ 0\,$, \eqref{solxxbot} yields
\begin{equation*}
  \sur{X}\,(\alpha)\ =\ \cos\,\left[\/\dbar\,\partial_{\,\alpha}\/\right]\,\bot{X}\,(\alpha)\,,
\end{equation*}
that can be inverted as
\begin{equation*}
  \bot{X}\,(\alpha)\ =\ \sec\,\left[\/\dbar\,\partial_{\,\alpha}\/\right]\,\sur{X}\,(\alpha)\,.
\end{equation*}
The relation \eqref{solyxbot} can then be rewritten with quantities expressed at the free surface only, \ie
\begin{equation}\label{relYXs}
  \sur{Y}\,(\alpha)\ =\ \mathscr{H}\,\left\{\,\sur{X}\,(\alpha)\,\right\}\,, \qquad 
  \mathscr{H}\ \eqdef\ \tan\,\left[\,\dbar\,\partial_{\,\alpha}\,\right]\,,
\end{equation}
where $\mathscr{H}$ is an anti-adjoint pseudo-differential operator acting on a 
pure frequency as
\begin{equation}\label{defHop}
  \mathscr{H}\!\left\{\ue^{\,\ui\,\kappa\,\alpha}\right\}\ =\ \dbar\,\frac{\partial\,\ue^{\,\ui\,\kappa\,\alpha}}{\partial\alpha}\ +\ \frac{\dbar^{\,3}}{3}\,\frac{\partial^{\,3}\,\ue^{\,\ui\,\kappa\,\alpha}}{\partial\alpha^{\,3}}\ +\ \frac{2\,\dbar^{\,5}}{15}\,\frac{\partial^{\,5}\,\ue^{\,\ui\,\kappa\,\alpha}}{\partial\alpha^{\,5}}\ +\ \cdots\ =\ \ui\,\tanh(\/\kappa\/\dbar\,)\,\ue^{\,\ui\,\kappa\,\alpha}\,.
\end{equation}
Indeed, the differential operator $\partial_{\,\alpha}$ corresponding to a 
simple multiplication by $\ui\,\kappa$ in \textsc{Fourier} space ($\kappa$ the frequency\footnote{$\kappa\ =\ n\,k$ for the $n$\up{th} \textsc{Fourier} mode of a ($2\pi/k$)-periodic function.}), the pseudo-differential operators are defined in \textsc{Fourier} space substituting $\ui\kappa$ for $\partial_{\,\alpha}\,$. Thus, for instance, the operator $\cos\,[(\beta\ +\ \dbar)\partial_{\,\alpha}]$ is in \textsc{Fourier} space a multiplication by $\cosh\,[\kappa\,(\beta\ +\ \dbar)]\,$. It should be noted that the formal \textsc{Taylor} expansions --- such as in \eqref{solxxbot} and \eqref{solyxbot} --- of the pseudo-differential operators are practically useless, specially in deep water ($\depth\ \to\ \infty$) where the operator $\mathscr{H}$ becomes the classical \textsc{Hilbert} transform.


\subsection{Averaging of dependent functions}

At the bottom, averaging $x$ over one wavelength, one obtains easily
\begin{equation}\label{avexal}
  \left\llangle\,\bot{x}\,\right\rrangle\ =\ \cl\,\ce^{\,-1}\left\llangle\,\alpha\,\right\rrangle\ +\ \left\llangle\,\bot{X}\,\right\rrangle\ =\ \upi\,k^{\,-1}\ +\ \left\llangle\,\bot{X}\,\right\rrangle\,.
\end{equation}
From the definition of the average operator \eqref{defmeanalpha}, we have also
\begin{equation}\label{avexal2}
  \left\llangle\,\bot{x}\,\right\rrangle\ =\ \left<\,-\,\ce^{\,-1}\,x\,\bot{\phi}_{\,x}\,\right>\,=\ \upi\,k^{\,-1}\,.
\end{equation}
Comparing \eqref{avexal} and \eqref{avexal2}, we obtain
\begin{equation}\label{aveX}
  \left\llangle\,\bot{X}\,\right\rrangle\ =\ 0\,,
\end{equation}
meaning that $\bot{X}$ is a periodic function averaging zero. Thus, with the relations \eqref{relYXs} to \eqref{aveX}, the boundedness and periodicity 
of $\bot{X}$ imply that
\begin{align}\label{meanY}
  \left\llangle\,X\,(\alpha,\,\beta)\,\right\rrangle\ =\ 0\,, \qquad
  \left\llangle\,Y\,(\alpha,\,\beta)\,\right\rrangle\ =\ 0\,.
\end{align}
Hence, the functions $X$ and $Y$ have zero average in the $\alpha-$variable. The relation \eqref{relYXs} can thus be inverted without ambiguities giving, in particular,
\begin{equation*}
  \sur{X}_{\,\alpha}\ =\ \mathscr{C}\!\left\{\sur{Y}\right\}\,, \qquad 
  \mathscr{C}\ \eqdef\ \partial_{\,\alpha}\cot\,\left[\,\dbar\,\partial_{\,\alpha}\,\right]\,.
\end{equation*}
$\mathscr{C}$ being a self-adjoint positive-definite pseudo-differential operator such that
\begin{equation}\label{defCop}
  \mathscr{C}\!\left\{\ue^{\,\ui\,\kappa\,\alpha}\right\}\ =\ \left\{
  \begin{array}{lr}
  \kappa\,\coth\,(\/\kappa\/\dbar\/)\,\ue^{\,\ui\,\kappa\,\alpha}  &\quad (\kappa\ \neq\ 0)\,, \\
   1\,/\,\dbar  &      \quad  (\kappa\ =\ 0)\,.
\end{array}
\right.
\end{equation}
Note that $\mathscr{C}_{\,\infty}^{\,-1}$ is singular for $\kappa\ =\ 0\,$, but the special choice $\mathscr{C}_{\,\infty}^{\,-1}\,\left\{1\right\}\ \eqdef\ 0$ does not matter as long as $\mathscr{C}_{\,\infty}^{\,-1}$ is applied to a function averaging to zero (or to any known value that can be enforced).

In summary, we have obtained the special relations
\begin{align}
\sur{x}_\alpha\ &=\ \cl\,\ce^{\,-1}\ +\ \mathscr{C}\!\left\{\sur{Y}\right\}\,=\ 
\cl\,\cs^{\,-1}\ +\ \mathscr{C}\!\left\{\sur{y}\right\}, \label{refxaeta} \\
\mathscr{C}\!\left\{\sur{Y}\right\}\,&=\ \mathscr{C}\!\left\{\sur{y}\right\}\ 
+\ \cl\,\cs^{\,-1}\ -\ \cl\,\ce^{\,-1}, 
\end{align}
and the averaged quantities.
\begin{align}\label{meanCY}
  \left\llangle\,\mathscr{C}\!\left\{\sur{Y}\right\}\,\right\rrangle\ =\ 0\,, \qquad
  \left\llangle\,\sur{y}\,\right\rrangle\ =\ \left(\,\cs\,\ce^{\,-1}\ -\ 1\,\right)\depth\,, \qquad
  \left\llangle\,\mathscr{C}\!\left\{\sur{y}\right\}\,\right\rrangle\ =\ \cl\,\ce^{\,-1}\ -\ \cl\,\cs^{\,-1}\,.
\end{align}


\subsection{Mean level condition}

The definition of the mean level in the transformed domain remains to be considered. Using the relations \eqref{defxyXY} and \eqref{meanCY}, the mean level condition \eqref{defmean} becomes
\begin{align*}
  0\ =\, \left\llangle\,\sur{y}\,\sur{x}_{\,\alpha}\,\right\rrangle\ =\ \left\llangle\left(\/\sur{Y}\/-\/\depth\/(1-\cs/\ce)\/\right)\!\left(\/\sur{X}_{\,\alpha}\ +\ \cl/\ce\/\right)\right\rrangle\ =\ \left\llangle\,\sur{Y}\,\sur{X}_{\,\alpha}\,\right\rrangle\ -\ \depth\left(\/1\/-\/\cs/\ce\/\right)\cl\,/\,\ce\,,
\end{align*}
thence
\begin{align}\label{minlevY}
  \left\llangle\,\sur{Y}\,\mathscr{C}\!\left\{\sur{Y}\right\}\,\right\rrangle\ &=\ \depth\,(\ce-\cs)\,\cl\,\ce^{\,-2}\ =\ -\,\cl\,\ce^{\,-1}\left\llangle\,\sur{y}\,\right\rrangle\,.
\end{align}
This relation is fundamental, in particular for a numerical resolution. With the previous results, several useful  averaged relations can be expressed in term of the important parameter $\left\llangle\sur{y}\right\rrangle$ (\cf Appendix~\ref{appaverel}).


\subsection{Celerities}

The definition \eqref{bernBxy} of the \textsc{Bernoulli} constant written at the free surface and at the bottom yields
\begin{equation}\label{defBCMbis}
  \frac{B}{\cl\,\ce}\ =\ \left\llangle\,\frac{\cl\,\cs^{\,-1}\ +\ \mathscr{C}\,\left\{\sur{y}\right\}}{\left(\cl\/\cs^{\,-1}\ +\ \mathscr{C}\,\left\{\sur{y}\right\}\right)^{\,2}\ +\ \sur{y}_{\,\alpha}^{\,2}}\,\right\rrangle\ =\ \left\llangle\,\frac{1}{\cl\,\cs^{\,-1}\ +\ \mathscr{S}\,\left\{\sur{y}\right\}}\,\right\rrangle\,,
\end{equation}
the second equality deriving from the relations
\begin{align*}
  \bot{u}\,/\,\cl\ =\ -\,1\,/\,\bot{x}_{\,\alpha}\,, \qquad
  \bot{x}_{\,\alpha}\ &=\ \sec\,\left[\/\dbar\,\partial_{\,\alpha}\/\right]\,\sur{x}_{\,\alpha}\ =\ \cl\,\cs^{\,-1}\ +\ \mathscr{S}\,\left\{\sur{y}\right\}\,,
\end{align*}
where $\mathscr{S}\ \eqdef\ \partial_{\,\alpha}\csc\,\left[\/\dbar\,\partial_{\,\alpha}\/\right]$ is a pseudo-differential operator acting on a pure frequency as
\begin{equation*}
  \mathscr{S}\,\left\{\ue^{\,\ui\,\kappa\,\alpha}\right\}\ =\ \left\{
  \begin{array}{lr}
    \kappa\,\operatorname{csch}\,(\/\kappa\/\dbar\/)\,\ue^{\,\ui\,\kappa\,\alpha}  & \quad  (\kappa\ \neq\ 0)\,,  \\
    1\,/\,\dbar  &      \quad  (\kappa\ =\ 0)\,.
  \end{array}
  \right. 
\end{equation*}
Note that, as $|\kappa|\ \to\ \infty\,$, $\mathscr{S}$ decays exponentially fast in \textsc{Fourier} space, unlike $\mathscr{C}$ that grows linearly. Note also that $\mathscr{S}\,\left\{\sur{y}\right\}\ \to\ 0$ as $\depth\ \to\ \infty\,$, hence $B\ =\ \ce\,\cs$ in deep water (together with $\ce\ =\ \cs\ =\ c$ as mentioned above) and thus $B\ =\ c^{\,2}\,$.


\section{Babenko equations}
\label{secbab}

Using \eqref{wuvqmap}({\it a}) and the relation $u^{\,2}\ +\ v^{\,2}\ =\ w\,w^*$, the \textsc{Bernoulli} equation \eqref{bernbase} at the free surface can be written
\begin{equation}\label{defwsur}
  \sur{w}\ =\ \frac{B\,-\,2\,g\,\eta}{\sur{w}^*}\ =\ \frac{2\,g\,\sur{y}\,-\,B}{\cl}\,\frac{\ud\,\sur{z}^*}{\ud\/\alpha}\ =\ \frac{(2\,g\,\sur{y}\,-\,B)\,\sur{x}_\alpha}{\cl}\ -\ \ui\,\frac{(2\,g\,\sur{y}\ -\ B)\,\sur{y}_\alpha}{\cl}\,.
\end{equation}
$w\ =\ u\ -\ \ui\,v$ being a holomorphic function such that $\Im\,(w)\ =\ 0$ at the bottom, we have at the free surface --- see the derivation of \eqref{relYXs} in the previous section ---
\begin{equation*}\label{relRSsur}
  -\,\sur{v}\,(\alpha)\ =\ \mathscr{H}\,\left\{\sur{u}\right\}\ =\ \tan\,[\/\dbar\,\partial_{\,\alpha}\/]\,\sur{u}\,(\alpha)\,,
\end{equation*}
thence, with \eqref{refxaeta} and \eqref{defwsur},
\begin{align*}
  {\partial_{\,\alpha}}\left({B\,\sur{y}\ -\ g\,\sur{y}^2}\right)\ &=\ \mathscr{H}\,\left\{\,(2\,g\,\sur{y}\ -\ B)\,\left(\cl\,\cs^{\,-1}\ +\ \mathscr{C}\,\left\{\sur{y}\right\}\right)\right\}\nonumber\\
  &=\ 2\,g\,\cl\,\cs^{\,-1}\,\mathscr{H}\,\left\{\sur{y}\right\}\ +\ 2\,g\,\mathscr{H}\,\left\{\sur{y}\,\mathscr{C}\!\left\{\sur{y}\right\}\right\}\ -\ B\,\sur{y}_{\,\alpha}\,.
\end{align*}
After simplifications and applying the antiderivative operator $\partial_{\,\alpha}^{\,-1}\,$, one obtains at once the \textsc{Babenko} equation
\begin{equation}\label{babeta}
  B\,g^{\,-1}\,\sur{y}\ -\ \half\,\sur{y}^{\,2}\ +\ K_{\,1}\ =\ \cl\,\cs^{\,-1}\,\mathscr{C}^{\,-1}\,\{\sur{y}\}\ +\ \mathscr{C}^{\,-1}\,\left\{\sur{y}\,\mathscr{C}\{\sur{y}\}\right\}\,,
\end{equation}
where $K_{\,1}$ is an integration constant. Averaging the equation over one wavelength, then using the relations \eqref{meanCieCM} and \eqref{meanCieCeCM}, one obtains an expression for the constant $K_{\,1}$
\begin{align}\label{K1}
  K_{\,1}\ &=\ \half\left\llangle\,\sur{y}^{\,2}\,\right\rrangle\ -\ B\,g^{\,-1}\left\llangle\,\sur{y}\,\right\rrangle\ \geqslant\ 0\,.
\end{align}

For numerical resolutions, it is convenient to make the change of dependent variable $\sur{y}\,(\alpha)\ \eqdef\ \Upsilon\,(\alpha)\ +\ \delta\,$, where $\delta$ is a constant at our disposal. Thus, the equation \eqref{babeta} becomes
\begin{equation}\label{babUp}
  \left(\/B\/g^{\,-1}\ -\ 2\/\delta\/\right)\Upsilon\ -\ \half\,\Upsilon^{\,2}\ +\ K_{\,2}\ =\ \left(\/1\ +\ \delta\/\depth^{\,-1}\/\right)\cl\,\cs^{\,-1}\,\mathscr{C}^{\,-1}\{\Upsilon\}\ +\ \mathscr{C}^{\,-1}\,\left\{\Upsilon\,\mathscr{C}\,\{\Upsilon\}\right\}\,,
\end{equation}
where
\begin{align}\label{defK2}
  K_{\,2}\ &=\ K_{\,1}\ -\ (\depth\ -\ B/g)\,\delta\ -\ \threehalf\;\delta^{\,2} \nonumber\\
  &=\ \left(\delta\ -\ B\,g^{\,-1}\right)\left\llangle\,\Upsilon\,\right\rrangle\ +\ \half\left\llangle\,\Upsilon^{\,2}\,\right\rrangle\ -\ (\depth\ +\ \delta)\,\delta\,,
\end{align}
and we have (see Appendix~\ref{appaverel} for details)
\begin{align*}
  \left\llangle\,\Upsilon\,\right\rrangle\ &=\ \left(\,\cs\,\ce^{\,-1}\ -\ 1\,\right)\depth\ -\ \delta\ =\ \left\llangle\,\sur{y}\,\right\rrangle\ -\ \delta\,,\\
  \left\llangle\,\Upsilon\,\mathscr{C}\,\left\{\Upsilon\right\}\,\right\rrangle\,&=\ -\,\cl\,\cs^{\,-1}\,(\,1\ +\ 2\,\delta\,\depth^{-1}\,)\left\llangle\,\Upsilon\,\right\rrangle\ -\ \cl\,\cs^{\,-1}\,(\,1\ +\ \delta\,\depth^{-1}\,)\,\delta\,,
\end{align*}
giving two equations for $\delta$ and $\cs\//\/\ce\,$. Alternatively, a related equation is obtained obtained applying the operator $\mathscr{C}$ to \eqref{babUp}
\begin{equation}\label{babUpbis}
  \left(\/B\/g^{-1}\ -\ 2\/\delta\/\right)\,\mathscr{C}\,\{\Upsilon\}\ -\ \half\,\mathscr{C}\,\left\{\Upsilon^2\right\}\ +\ K_{\,2}\,{\dbar}^{\,-1}\ =\ \left(\/1\ +\ \delta\/\depth^{\,-1}\/\right)\cl\,\cs^{\,-1}\,\Upsilon\ +\ \Upsilon\,\mathscr{C}\,\{\Upsilon\}\,.
\end{equation}

In practice, the parameter $\delta$ may be conveniently chosen such that:
\begin{itemize}
  \item[(i)] $\delta\ =\ 0\,$, so we are dealing with the free surface $\sur{y}\,$;
  \item[(ii)] $\delta\ =\ \left\llangle\,\sur{y}\,\right\rrangle\,$, so we are dealing with the zero-mean dependant variable $\sur{Y}\,$; 
  \item[(iii)] $\delta\ =\ \min\,(\sur{y})\,$, so $\Upsilon\ \geqslant\ 0$ and the wave appears to somehow look like a ``solitary wave'' on $[\,-\pi/k,\,\pi/k\,]\,$; 
  \item[(iv)] $K_{\,2}\ =\ 0\,$, so the equation \eqref{babUp} is homogeneous. 
\end{itemize}
Properties (iii) and (iv) are both desirable features for using \textsc{Petviashvili}'s iterations, leading to a simple and fast numerical scheme. However, no choice of $\delta$ can enforce (iii) and (iv) simultaneously, so we proceed as follow.

Note first that in deep water $\depth\ \to\ \infty\,$, $\mathscr{C}^{\,-1}\ \to\ \mathscr{C}_{\,\infty}^{\,-1}$ is singular and the equation \eqref{babUp} is problematic because $\mathscr{C}_{\,\infty}^{-1}$ is applied to the quantity $\Upsilon\,\mathscr{C}_{\,\infty}\,\left\{\Upsilon\right\}$ that has a non-zero (unknown a priori) mean value. This problem is overcome applying the operator $\mathscr{C}_{\,\infty}$ to \eqref{babUp} (the constant $K_{\,2}$ then vanishes), thus yielding the alternative homogeneous equation
\begin{multline}\label{babYter}
  \left(\frac{B}{g}\ -\ 2\,\delta\,\right)\,\mathscr{C}_{\,\infty}\,\left\{\Upsilon\right\}\ -\ \mathscr{C}_\infty\!\left\{\frac{\Upsilon^{\,2}}{2}\right\}\ =\\ 
  \left(1\ +\ \frac{\delta}{\depth}\right)\frac{\cl}{\cs}\,\mathscr{C}_{\,\infty}\,\circ\,\mathscr{C}^{\,-1}\,\left\{\Upsilon\right\}\ +\ \mathscr{C}_{\,\infty}\,\circ\,\mathscr{C}^{\,-1}\,\left\{\Upsilon\,\mathscr{C}\,\left\{\Upsilon\right\}\right\}\,,
\end{multline}
where
\begin{equation*}
  \mathscr{C}_{\,\infty}\circ\mathscr{C}^{\,-1}\,\left\{\ue^{\,\ui\,\kappa\,\alpha}\right\}\ =\ \tanh\,|\/\kappa\/\dbar\,|\;\ue^{\,\ui\,\kappa\,\alpha}\,.
\end{equation*}
The equation \eqref{babYter} for deep water with $\delta\ =\ 0$ was solved by \cite{Dyachenko2014} using the generalised \textsc{Petviashvili} method (GPM) because the classical \textsc{Petviashvili} method (CPM) does not converge (with $\delta\ =\ 0$). Here, we solve the equation \eqref{babYter} with $\delta\ =\ \min\,(\tilde{y})$ using classical \textsc{Petviashvili}'s iterations that converge in deep water and in finite depth.


\section{Numerical resolution}
\label{secnum}

We detail here how the numerical algorithm is implemented. We mostly focus on periodic waves ($k\ >\ 0$), the case of solitary waves ($k\ =\ 0$) being subsequently briefly described. Details for alternative computations of solitary waves are given in \cite{Clamond2012b, Dutykh2013b}.


\subsection{User provided parameters}
\label{ssecusepar}

For practical applications, the user generally wants to define the wave by the parameters $g\,$, $\depth\,$, $k$ and $H\,$. Two of these parameters can be freely chosen to compute the solution. In deep water and finite depth (\ie not shallow), a wave is most often defined only by the two dimensionless parameters $k\depth$ and $\varepsilon\ =\ kH/2\,$, from which the four physical parameters are defined as follow in the numerical resolution. In shallow water, the dimensionless height $H/d$ is often used instead of $\varepsilon\,$. It should be noticed that $\varepsilon$ is not a good parameter to characterise the steepness in shallow water ($\varepsilon\ =\ 0$ for all solitary waves), while $H/d$ is not suitable in deep water ($H/d\ =\ 0$ for all waves in infinite depth). Alternative definitions of the wave steepness, valid for all depths and all wavelengths, could be introduced such as $kH/\tanh(kd)\,$. However, this alternative steepness parameter being generally not used by practitioners, it is not considered here.

The water depth is considered infinite if 
\begin{equation}\label{defdeep}
  1\ -\ \tanh(k\depth)\ \leqslant\ \mathsf{tol},
\end{equation}
where $\mathsf{tol}$ is the numerical tolerance used for the computations. Working with $n-$digit arithmetics, we generally take $\mathsf{tol}\ =\ 10^{\,2\,-\,n}$ yielding $\mathsf{tol}\ =\ 10^{\,-14}$ in double precision, $\mathsf{tol}\ =\ 10^{\,-32}$ in quadruple precision and $\mathsf{tol}\ =\ 10^{\,-69}$ in octuple precision (standard  IEEE 754--2008). If the inequality \eqref{defdeep} is fulfilled, we consider that $\depth\ =\ \infty$ and we choose (dimensionless) units such that $g\ =\ k\ =\ 1$  (this choice is always possible via a suitable scaling). Otherwise, we choose $g\ =\ \depth\ =\ 1$ (that is also always possible without loss of generality). Many other scaling could be used, but the ones above are the most common in both deep water and finite depth.

The parameters $g\,$, $\depth$ and $k$ being now defined, and the parameter $\varepsilon$ being chosen, the total wave height is obviously $H\ =\ 2\varepsilon/k\,$. Note that the wave can be defined by other pairs of dimensionless parameters than $\{kd,\,\varepsilon\}$ or $\{kd,\,H/d\}\,$. For instance, a parameter like $|\tilde{u}_{\,0}/\cl|$ ($\tilde{u}_{\,0}$ the horizontal velocity at the crest) is often used to investigate the (almost) highest waves 
\cite{Maklakov2002, Maklakov2015}. However, this parameter involves quantities generally unknown from, \eg, experimental measurements, so it is not convenient for a practical use. Thus, defining a wave by such parameters would require important pre- and post-processing for practical applications, that are not always trivial.


\subsection{Computational parameters}
\label{sseccompar}

In addition to the physical parameters defined above, it is convenient (but unessential) to take $\cl\ =\ \ce$ and to introduce the dimensionless parameter $\sigma\ \eqdef\ \cs/\ce$ (hence $\kbar\ =\ k\,$, $\dbar\ =\ \sigma\depth$). In infinite depth $\sigma\ =\ 1$, but in finite depth $\sigma$ is unknown and therefore must be computed (see below). We also chose $\delta\ =\ \min(\sur{y})$ because we shall use \textsc{Petviashvili}'s iterations.

The fundamental period is discretised with $2\,N$ equally spaced points $\kbar\/\alpha_{\,j}\ =\ (j\ -\ 1)\pi/N$ for $j\ =\ 1,\,2,\,\ldots,\,2\,N\,$. The number of nodes should be large enough to ensure the desired accuracy. For infinitesimal short waves (\ie $\varepsilon\ \ll\ 1$ and $k\,d$ not small), $N\ =\ 32$ (say) may be sufficient to ensure machine precision when working in double precision. However, the number of required nodes increases rapidly as the wave-height and the wavelength increase. Thus, for steep and cnoidal waves, the number of nodes must be substantially larger. Since we are solving the equation with an algorithm requiring $\O\,(N\log N)$ operations, $N$ can be large without significant speed burden. Therefore, one can take $N\ =\ 1\,024$ as default when working in double precision because it is often sufficient. If more nodes are needed, the cost of increasing $N$ is low because the algorithm is very fast. However, $N$ cannot be too large in order to avoid large accumulation of round-off errors (when working in double precision, it is wise to take $N\ \leqslant\ 2^{\,19}\ =\ 524\,288\,$, say). If an extremely large number of nodes is required, then quadruple or higher precision is necessary to avoid a significant accumulation of round-off errors.


\subsection{Classical Petviashvili method (CPM)}
\label{ssecpetite}

\textsc{Petviashvili}'s iterations are a simple and efficient algorithm for computing solitary waves solution of the \textsc{Babenko} equation \cite{Clamond2012b, Dutykh2013b}. This algorithm is very easy to implement and runs fast since each iteration requires $\O\,(N)$ operations. Note that \textsc{Petviashvili}'s iterations are used here together with fast \textsc{Fourier} transforms (FFT) that require $\O\,(N\log N)$ operations, so the overall complexity of our algorithm is $N\log N\,$.

In order to apply \textsc{Petviashvili}'s method, we separate the linear and nonlinear terms and rewrite the equation \eqref{babYter} as $\mathscr{L}\,\{\Upsilon\}\ =\ \mathscr{N}\,\{\Upsilon\}$ with
\begin{align}
  \mathscr{L}\{\Upsilon\}\ &\eqdef\ \left(B/g\,-\,2\,\delta\,\right)\mathscr{C}_\infty\!\left\{\Upsilon\right\}\ -\ \left(1\,+\,\delta/\depth\right) \sigma^{\,-1}\,\mathscr{C}_\infty\!\circ\!\mathscr{C}^{-1}\!\left\{\Upsilon\right\}, 
\label{defLUps} \\
  \qquad \mathscr{N}\{\Upsilon\}\ &\eqdef\ \half\,\mathscr{C}_\infty\!\left\{\Upsilon^2\right\}\ +\ \mathscr{C}_\infty\!\circ\!\mathscr{C}^{\,-1}\!\left\{\Upsilon\,\mathscr{C}\!\left\{\Upsilon\right\}\right\}, \label{defNUps}
\end{align}
where $\sigma\ =\ \cs/\ce\,$, $\delta\ =\ \min(\sur{y})$ and $\Upsilon\ =\ y\ -\ \delta\,$. From the approximation $\Upsilon^{\,(i)}$ obtained at the $i$\up{th} iterations, the next approximation $\Upsilon^{\,(i+1)}$ is obtained via the \textsc{Petviashvili} weighted fixed point iteration
\begin{equation}\label{petite}
  \Upsilon^{\,(i+1)}\ =\ S_{\,i}^{\,2}\times\mathscr{L}^{\,-1}\!\circ\mathscr{N}\!\left\{\Upsilon^{(i)}\right\}, \qquad
  S_{\,i}\ \eqdef\ \left.{\left\llangle\,\Upsilon^{\,(i)}\,\mathscr{L}\!\left\{\Upsilon^{\,(i)}\,\right\}\right\rrangle}\,\right/\,{\left\llangle\,\Upsilon^{(i)}\,\mathscr{N}\!\left\{\Upsilon^{\,(i)}\right\}\,\right\rrangle}.
\end{equation}
At each iteration, the zero value of $\Upsilon^{\,(i)}$ at the trough and the wave height are enforced via the renormalisation
\begin{equation}\label{renUps}
  \Upsilon^{\,(i+1)}(\alpha)\ \longleftarrow\ H\,\frac{\Upsilon^{\,(i+1)}(\alpha)\,-\,\Upsilon^{\,(i+1)}(\kbar/\pi)}{\Upsilon^{\,(i+1)}(0)\,-\,\Upsilon^{\,(i+1)}(\kbar/\pi)}\,.
\end{equation}
This renormalisation improves the convergence, specially for steep waves.

In finite depth, the operator $\mathscr{L}^{\,-1}$ is singular but $\mathscr{L}^{\,-1}\circ\mathscr{N}$ is regular. This is because both operators involve the factor $\mathscr{C}_\infty$ that is zero for the zero frequency (this factor is introduced to kill the constant $K_{\,2}$). Doing so, the \textsc{Babenko} equation has been singularised at the zero frequency, but this singularity is only apparent (\ie movable in the sense that $\mathscr{C}_\infty^{\,-1}\circ\mathscr{C}_\infty$ is identity). However, defining explicitly $\mathscr{L}^{\,-1}\circ\mathscr{N}$ at the zero frequency is not necessary, the mean value of $\Upsilon^{\,(i+1)}$ being enforced by the renormalisation \eqref{renUps}. Indeed, setting arbitrarily $\mathscr{L}^{\,-1}\,\{1\}\ \equiv\ 0\,$, $\Upsilon^{\,(i+1)}$ computed with \eqref{petite} is obtained modulo an unknown (generally incorrect) mean value, the right value of $\left\llangle\Upsilon\right\rrangle$ being subsequently enforced via \eqref{renUps}.

As initial guess $\Upsilon^{\,(0)}\,$, we take the linear approximation
\begin{equation}\label{inigues}
  \Upsilon^{(0)}\ =\ \left[\,1\,+\,\cos(\kbar\alpha)\,\right]H\,/\,2, \qquad \sigma^{(0)}\ =\ 1,
\end{equation}
unless a better guess is provided by the user, for instance from another calculation with slightly different parameters $k\depth$ and $\varepsilon$ (useful for analytic continuations). We found that, with the initial guess \eqref{inigues}, the \textsc{Petviashvili} iterations always converge, even for large waves in shallow water, so we did not try out other guesses. The convergence from the initial guess \eqref{inigues} illustrates the robustness of the method.

\textsc{Petviasvili}'s iterations \eqref{petite} involve the unknown parameters $B\,$, $\sigma$ and $\delta$ via the definition of the operators $\mathscr{C}\,$, $\mathscr{L}$ and $\mathscr{N}\,$. Therefore, $B\,$, $\sigma$ and $\delta$ must be computed from $\Upsilon^{\,(i)}$ before \eqref{petite} can be used. (It would be the same with any other iterations, such as \textsc{Newton} and \textsc{Levenberg--Marquardt} methods.) These parameters are obtained as follow.


\subsection{Computation of the unknown parameters}

We first compute $\sur{Y}\ =\ \Upsilon^{\,(i)}\ -\ \left\llangle\,\Upsilon^{\,(i)}\,\right\rrangle\,$. In deep water $\sigma\ =\ 1$ and, according to \eqref{minlevY}, we have 
\begin{equation} \label{my0deep}
  \left\llangle\,\sur{y}\,\right\rrangle\ =\ -\left\llangle\,\sur{Y}\,\mathscr{C}_\infty\!\left\{\sur{Y}\right\}\,\right\rrangle,
\end{equation}
thence $\sur{y}\ =\ \sur{Y}\ +\ \left\llangle\,\sur{y}\,\right\rrangle$ by definition of $\sur{Y}\,$.

In finite depth, in general $\sigma\ \neq\ 1$ is unknown ($\sigma\ =\ 1$ only for solitary waves) and must be computed. To do so, the relation \eqref{minlevY} is rewritten as the equation
\begin{align}\label{eqsigma}
  E\,(\sigma)\ \eqdef\ \left\llangle\,\sur{Y}\,\mathscr{C}\!\left\{\sur{Y}\right\}\,\right\rrangle\ +\ (\sigma-1)\,\depth\ =\ 0\,.
\end{align}
It should be recalled here that $\mathscr{C}$ depends on $\sigma\,$, so \eqref{eqsigma} is a nonlinear equation for $\sigma\,$. Equation \eqref{eqsigma} is thus solved with \textsc{Newton} iterations
\begin{equation}\label{calcsigma}
  \sigma_{\,j+1}\ =\ \sigma_{\,j}\ -\ \frac{E(\sigma_j)}{E'(\sigma_j)}, \qquad 
  E^{\,\prime}\,(\sigma)\ \eqdef\ \frac{\ud\,E(\sigma)}{\ud\/\sigma}\ =\ \depth\ -\ \depth\left\llangle\,\sur{Y}\,\mathscr{S}^{\,2}\!\left\{\sur{Y}\right\}\,\right\rrangle\,,
\end{equation}
with $\mathscr{S}\ =\ \partial_\alpha\csc\!\left[\/\sigma\/\depth\/\partial_\alpha\/\right]\,$. In practice, one \textsc{Newton} iteration is sufficient because the initial guess $\sigma_{\,0}$ is given by the approximation of $\sigma$ obtained at the previous iteration from $\Upsilon^{\,(i-1)}$ that, if the CPM converges, is closed to the exact solution. Once $\sigma$ has been obtained, we compute $\left\llangle\sur{y}\right\rrangle\ =\ (\sigma\,-\,1)\,d$ and $\delta\ =\ \left\llangle\sur{y}\right\rrangle\ -\ \left\llangle\Upsilon\right\rrangle\,$, so these parameters are now known for the $i$\up{th} \textsc{Petviashvili} iteration. It should be emphasised that accurate computations of $\sigma$ and $\left\llangle\sur{y}\right\rrangle$ are absolutely crucial to ensure that the mapping $z\ \mapsto\ \zeta$ is conformal and that the still water level is where it should be.

Finally, the \textsc{Bernoulli} constant $B$ is obtained from the equation \eqref{babUpbis} applied at the crest ($\alpha\ =\ 0$) and at the trough ($\alpha\ =\ \pi/\kbar$), \ie,
\begin{align}
  K_2\ &=\ \left(\/\depth\/+\/\delta\/\right)H\ +\ \sigma\,\depth\,\left[\,\left(H+2\delta-B/g\right)\mathscr{C}\{\Upsilon\}\,+\,\half\,\mathscr{C}\{\Upsilon^2\}\,\right]_{0}\,,\label{defK22} \\
  &=\ \sigma\,\depth\,\left[\,\left(2\delta-B/g\right)\mathscr{C}\{\Upsilon\}\,+\,\half\,\mathscr{C}\{\Upsilon^2\}\,\right]_{\,\pi/\kbar}\,,\label{defK23}
\end{align}
thence
\begin{align}\label{solB}
  \frac{B}{g}\ =\ 2\,\delta\ -\ \frac{1\,+\,(\delta/\depth)\,+\,\sigma\left[\/\mathscr{C}\{\Upsilon\}\/\right]_{\,0}}{\sigma\left[\/\mathscr{C}\{\Upsilon\}\/\right]_{\,0}^{\,\pi/\kbar}}\,H\ +\ \frac{\left[\/\mathscr{C}\{\Upsilon^2\}\/\right]_{0}^{\pi/\kbar}}{2\left[\/\mathscr{C}\{\Upsilon\}\/\right]_{\,0}^{\,\pi/\kbar}},
\end{align}
with the notations $[f]_{\,a}\ \eqdef\ f(a)$ and $[f]_{\,a}^{\,b}\ \eqdef\ f\,(b)\ -\ f\,(a)\,$. Relation \eqref{solB} is obtained subtracting \eqref{defK22} and \eqref{defK23}, thus $K_{\,2}$ vanishes and does not need to be computed.

All the parameters involved in the \textsc{Babenko} equation are now defined and the \textsc{Petviashvili} iterations \eqref{petite} can be applied until the desired accuracy is reached.


\subsection{Post processing}

After convergence of the \textsc{Petviashvili} iterations, all the parameters of interest can be computed. The celerity $\ce$ is given by 
\begin{equation}\label{defberndag}
  \ce\ =\ \sqrt{B}\,\left\llangle\,\frac{1\,+\,\mathscr{C}\!\left\{\sur{Y}\right\}}{\left(1+\mathscr{C}\!\left\{\sur{Y}\right\}\right)^2\/+\,\sur{Y}_\alpha^{\,2}}\,\right\rrangle^{-1/2}
\end{equation}
thence $\cs\ =\ \sigma\ce\,$. These parameters being defined, all the integral quantities in Appendix~\ref{appint} are easily computed.

Often, users want to know the velocity and other fields inside the bulk at a given location $z\,$. This can be obtained from the integrals provided in the Appendix~\ref{appvel} and discretised according to the trapezoidal rule \cite{Dutykh2013b}. This is very simple to implement and also very accurate, provided that $z$ is not too close to the free surface. (Typically, the distance between $z$ and the free surface should be larger than $\Delta\alpha\,$.)

\begin{remark}
Since we use a spectral method to solve a nonlinear pseudo-differential equation, some aliasing errors may occur. With a number of \textsc{Fourier} modes large enough to achieve spectral accuracy, the magnitude of all the neglected high-frequency modes is smaller than the machine precision (about $10^{\,-16}$ in double precision). The equation being quadratic in nonlinearity, in theory a two-thirds rule \citep{Boyd2000} should be applied in order to avoid aliasing. However, even if the two-thirds rule is not applied, the aliasing error is about machine precision, provided that the aliased frequencies (\ie the upper third of the spectra) are small (say of magnitudes of at most $10^{\,-8}$ in double precision) so that their products are numerically zero. This means that an anti-aliasing filter is not necessary if $N$ is large enough, as can be seen in the spectra below. Of course, one can easily introduce an anti-aliasing (\ie low-pass) filter if this turns out to be necessary for some computations.
\end{remark}


\section{Numerical examples}
\label{secexa}

The algorithm described above has been implemented in Matlab\textsuperscript{\tiny{TM}} and it is freely downloadable \cite{Clamond2017d}. This program was written with clarity in mind, so it can be easily understood, modified and translated into any programming language. In particular, this program can be easily modified to run in arbitrary precision, provided that this feature is available to the user (we use the Advanpix Multiprecision Toolbox \cite{MATLAB2012}).

\begin{figure}
  \centering
  \vspace*{0.80em}
  \includegraphics[width=0.99\textwidth]{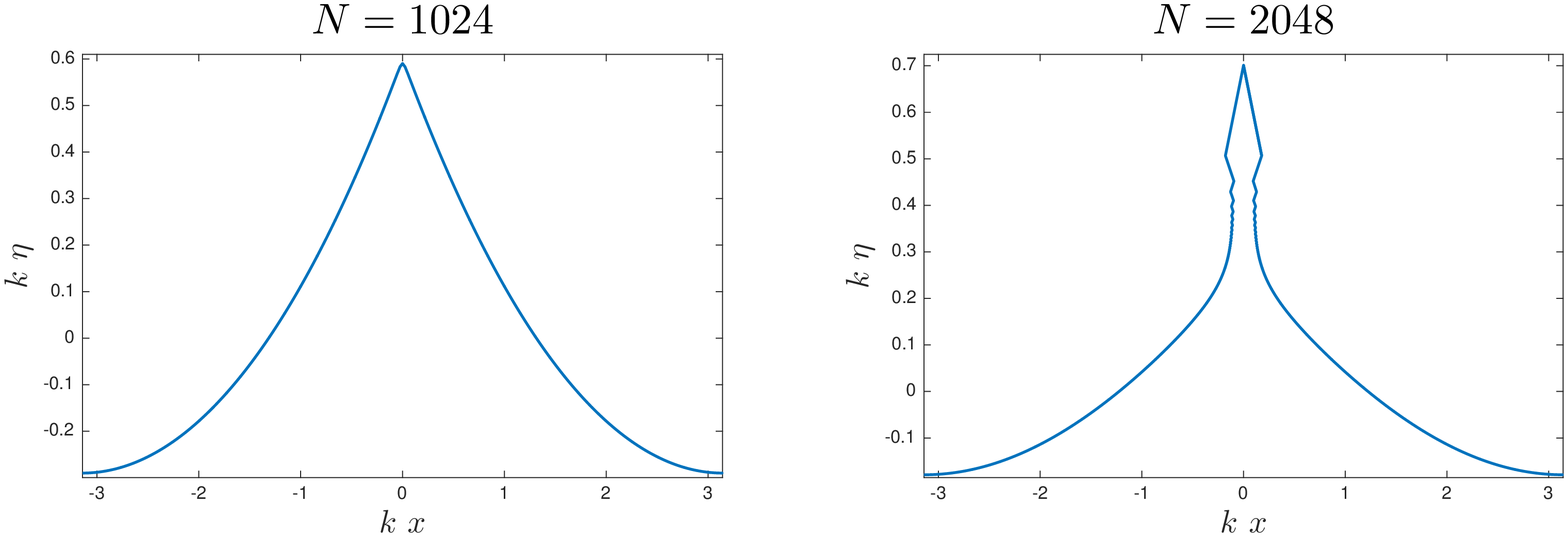}
  \caption{\small\em Influence of the number of \textsc{Fourier} modes for $kd\ =\ \infty\,$, $\varepsilon\ =\ 0.4401\,$.}
  \centerline{\scriptsize Left: normal solution obtained with $N\ =\ 1\,024\,$. Right: ghost solution obtained with $N\ =\ 2\,048\,$.}
  \label{figdeep4401}
\end{figure}

\begin{figure}
  \centering
  \vspace*{0.80em}
  \includegraphics[width=0.99\textwidth]{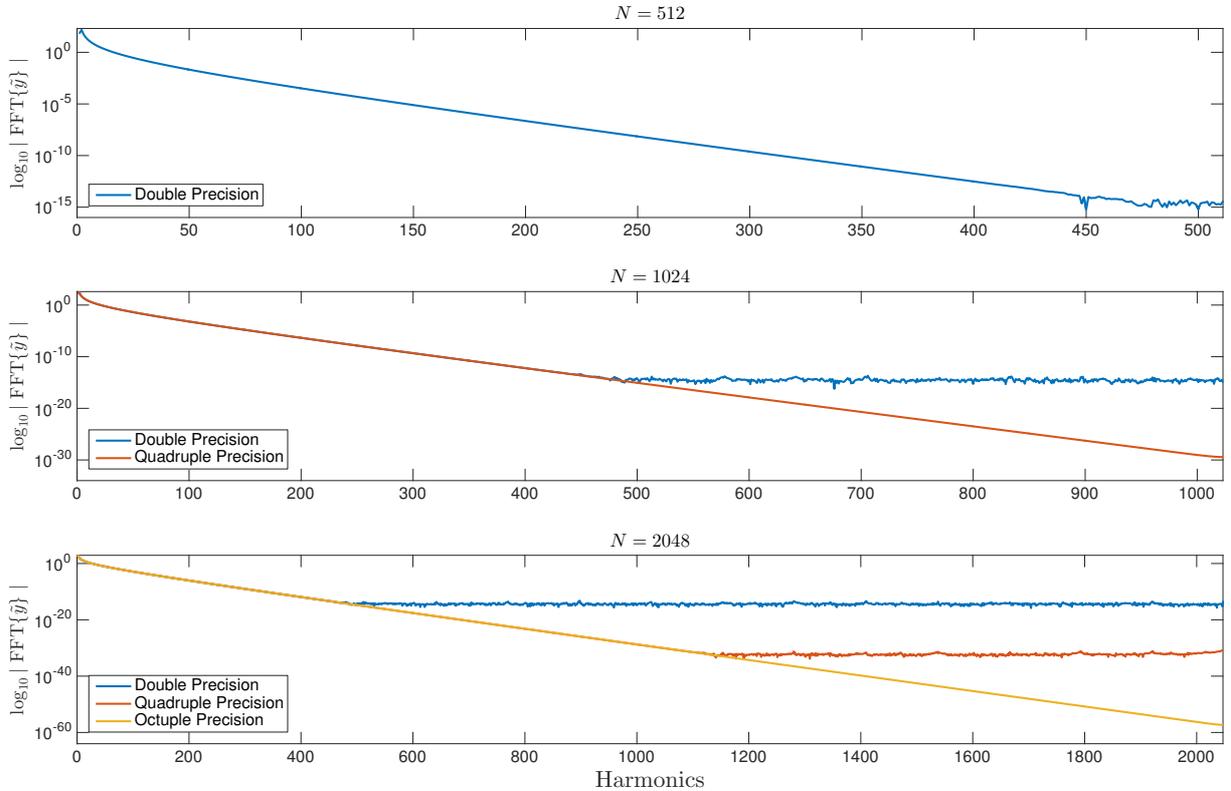}
  \caption{\small\em Decay of the \textsc{Fourier} coefficients for $\varepsilon\ =\ 0.4$ and $kd\ =\ \infty\,$.}
  \centerline{\scriptsize Blue: $16$ digits; Red: $34$ digits; Orange: $71$ digits.}
  \label{figdeepspec}
\end{figure}


\subsection{Deep water}

In deep water ($d\ \to\ \infty$), periodic waves with identical crests are obtained with our algorithm provided that $\varepsilon\ \lesssim\ 0.44\,$. Thus, the algorithm converges for rather steep waves (up to about $99.3\%$ of the highest waves), the maximum steepness being $\varepsilon\ \approx\ 0.443164$ \cite{Maklakov2002}.

It should be noted that with $\delta\ =\ 0$ the CPM diverges for all steepnesses. This shows that the choice $\delta\ =\ \min(\tilde{y})$ improves significantly the convergence of the CPM, rending unnecessary the use of the GPM. It should also be noted that \textsc{Fenton}'s algorithm \cite{Fenton1988} converges for $\varepsilon\ \lesssim\ 0.36$ and that it is much slower than the method described here. For $\varepsilon\ \lesssim\ 0.36\,$, \textsc{Fenton}'s and ours algorithms match up to about six digits (for non-infinitesimal waves), that corresponds to the accuracy of \textsc{Fenton}'s algorithm.

Actually, the present algorithm can converge also for $\varepsilon\ >\ 0.44\,$, but then `ghost' solutions \cite{Domokos2003} are obtained (similar to the one on the right Figure~\ref{figdeep4401}). For $\varepsilon\ =\ 0.44\,$, varying $N$ leads to the same solution in double and quadruple precisions. However, for $\varepsilon\ =\ 0.4401\,$, with $N\ =\ 1\,024$ a `normal' solution is obtained (Figure~\ref{figdeep4401} left), but a `ghost' (spurious) solution is obtained with $N\ =\ 2\,048$ (Figure~\ref{figdeep4401} right). The algorithm behaviour for the highest computable waves is discussed in the Section~\ref{sseclim} below.

For $\varepsilon\ \leqslant\ 0.44$ the algorithm converges rapidly\footnote{On a 2012 MacBook Pro laptop computer, with $N\ =\ 2\,048\,$, $\mathsf{tol}\ =\ 10^{\,-15}\,$, $\varepsilon\ =\ 0.4$ and in double precision, the solution is obtained in about half a second.} to the solution. Actually, any arbitrary accuracy can be achieved provided that $N$ is large enough (Figure~\ref{figdeepspec}). For instance, for $\varepsilon\ =\ 0.4\,$, the solution is obtained to machine double precision with $N\ =\ 512$ (Fig.~\ref{figdeepspec} upper). $N\ =\ 1\,024$ is not sufficient to achieve machine quadruple precision (Fig.~\ref{figdeepspec} middle), the latter being obtained for $N\ =\ 2\,048$ (Fig.~\ref{figdeepspec} lower). However, $N\ =\ 2\,048$ is not sufficient to achieve full octuple precision, that can be obtained with larger $N\,$. Similarly, any accuracy can be obtained provided that $N$ is large enough. Note that the Figure~\ref{figdeepspec} clearly shows that aliasing errors are not significant although no special anti-aliasing techniques were applied.

This test illustrates the accuracy and the robustness of the algorithm. Indeed, some algorithms diverge when $N$ is too large although they converge for smaller $N$ (for the same steepness); a problem not faced by the algorithm described here.

\begin{figure}
  \centering
  \vspace*{0.80em}
  \includegraphics[width=0.99\textwidth]{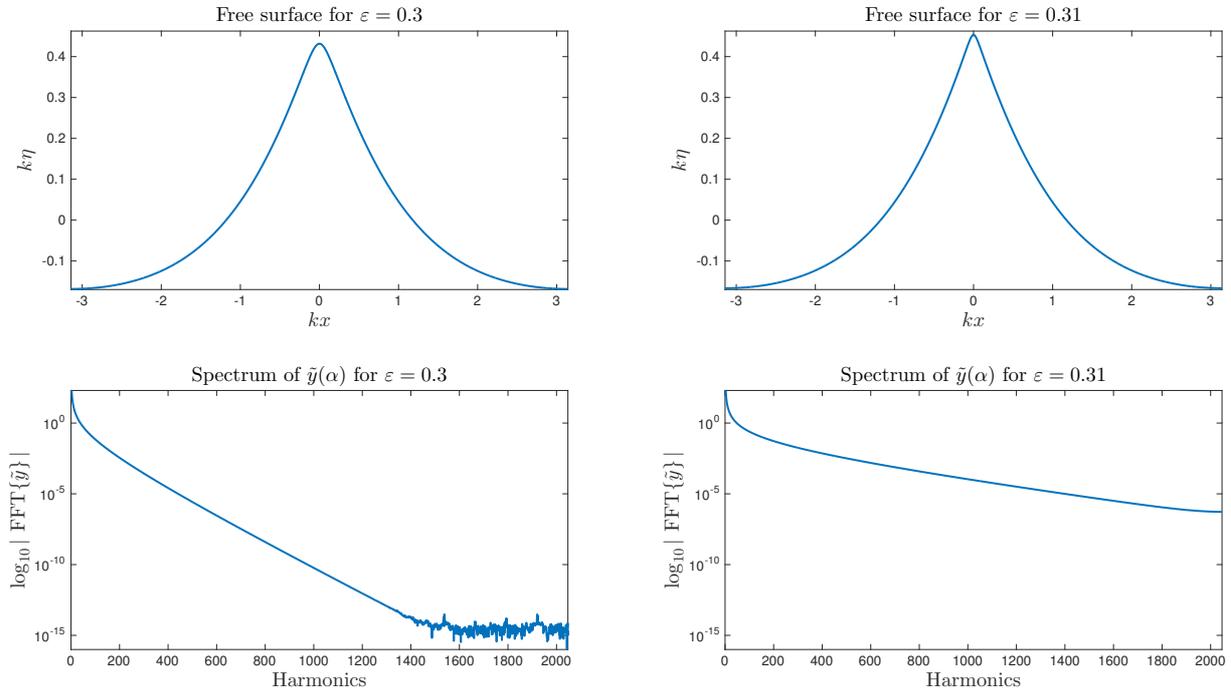}
  \caption{\small\em Examples of steep waves in finite depth for $kd\ =\ 1\,$.}
  \centerline{\scriptsize Left: $\varepsilon\ =\ 0.3$; Right: $\varepsilon\ =\ 0.31\,$; Upper: free surfaces; Lower: spectra.}
  \label{figfinisurf}
\end{figure}


\subsection{Finite depth}

We found that the CPM and GPM are both divergent when applied to the \textsc{Babenko} equation \eqref{babYter} with $\delta\ =\ 0\,$. Conversely, the CPM converges well if one takes $\delta\ =\ \min(\tilde{y})\,$, so we did not try the GPM. As for the deep water case, the algorithm converges for all but the highest waves, \ie, the algorithm converges up to the first maximum of $B\,(\varepsilon)\,$. For higher waves, the algorithm diverges or converges to a ghost solution, the maximum steepness computable depending on the depth (see section~\ref{sseclim}).

For example, consider the case $kd\ =\ 1$ with steepnesses $\varepsilon\ =\ 0.3$ and $\varepsilon\ =\ 0.31\,$. Although these two large steepnesses are close, they correspond to quite different free surfaces (Fig.~\ref{figfinisurf} upper), as can be seen in their spectra (Fig.~\ref{figfinisurf} lower). With $N\ =\ 2\,048$, the case $\varepsilon\ =\ 0.3$ is resolved to machine double precision (Fig.~\ref{figfinisurf} lower left), while the case $\varepsilon\ =\ 0.31$ is resolved only to a mild accuracy (Fig.~\ref{figfinisurf} lower right). Machine double precision is achieved for $\varepsilon\ =\ 0.31$ with $N\ =\ 8\,192\,$, however. This shows that as the wave approches the highest one, the number $N$ of \textsc{Fourier} modes has to be increased dramatically (exponentially fast) in order to reach a full spectral precision. This is inherent to the solution formulated in the conformal plane and it has nothing to do with the numerical algorithm described in the present paper. Indeed, for the highest wave with an angular crest, the \textsc{Fourier} spectrum decays algebraically while smaller waves have spectra decaying exponentially fast at high frequencies.

This example illustrates, like the one in deep water, the need for a large number of \textsc{Fourier} modes in order to achieve the full precision of a given floating point format. This example also illustrates the rapid increases of $N$ necessary for an accurate resolution as the steepness increases and, therefore, the need for a fast algorithm. Since the algorithm described here has an overall complexity $\O\,(N\log N)\,$, the necessity of large $N$ is not problematic and accurate computations are rapidly achieved.

For $kd\ =\ 1$ the highest computable wave has steepness $\varepsilon\ \approx\ 0.3146\,$, that is approximately $99.6\%$ of the highest wave. Limiting values for other $kd$ are given in the Table~\ref{tabHigh} showing that the highest computable wave, for any given depth and wavelength, is about $99\%$ of the maximum one.

\begin{figure}
  \centering
  \vspace*{0.80em}
  \includegraphics[width=0.99\textwidth]{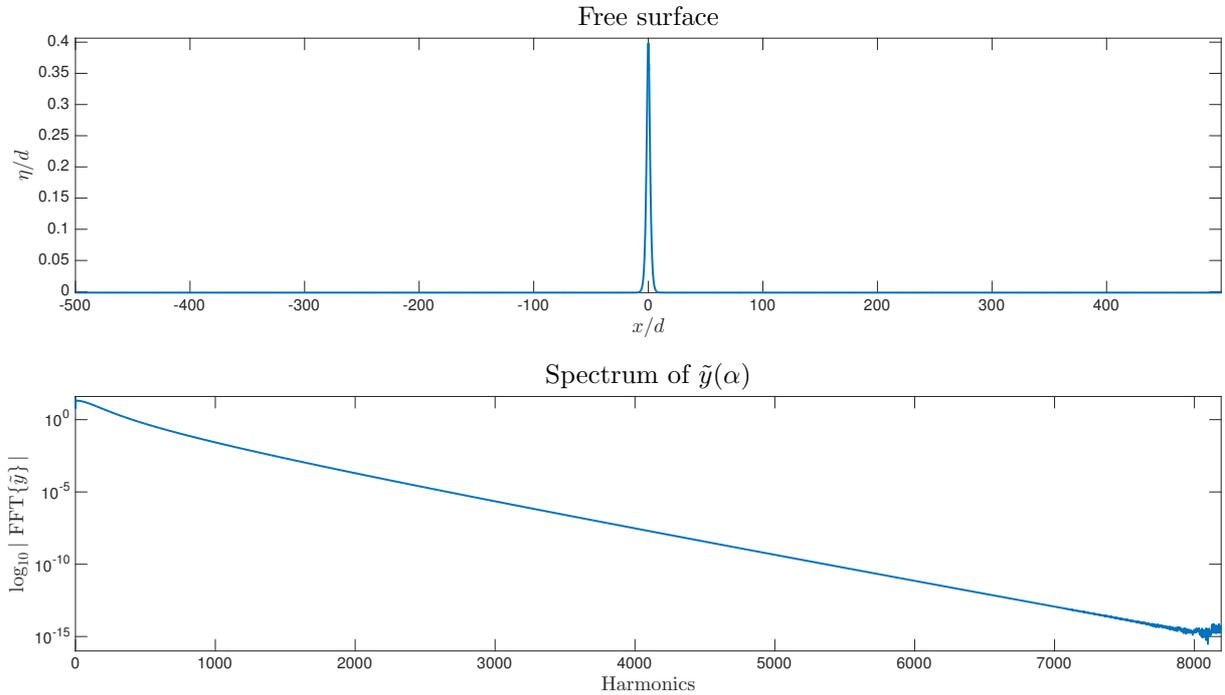}
  \caption{\small\em Cnoidal wave in very shallow water ($L/d\ =\ 1000\,$, $H/d\ =\ 0.4$).}
  \label{figcnoisurf}
\end{figure}

\begin{figure}
  \centering
  \vspace*{0.80em}
  \includegraphics[width=0.99\textwidth]{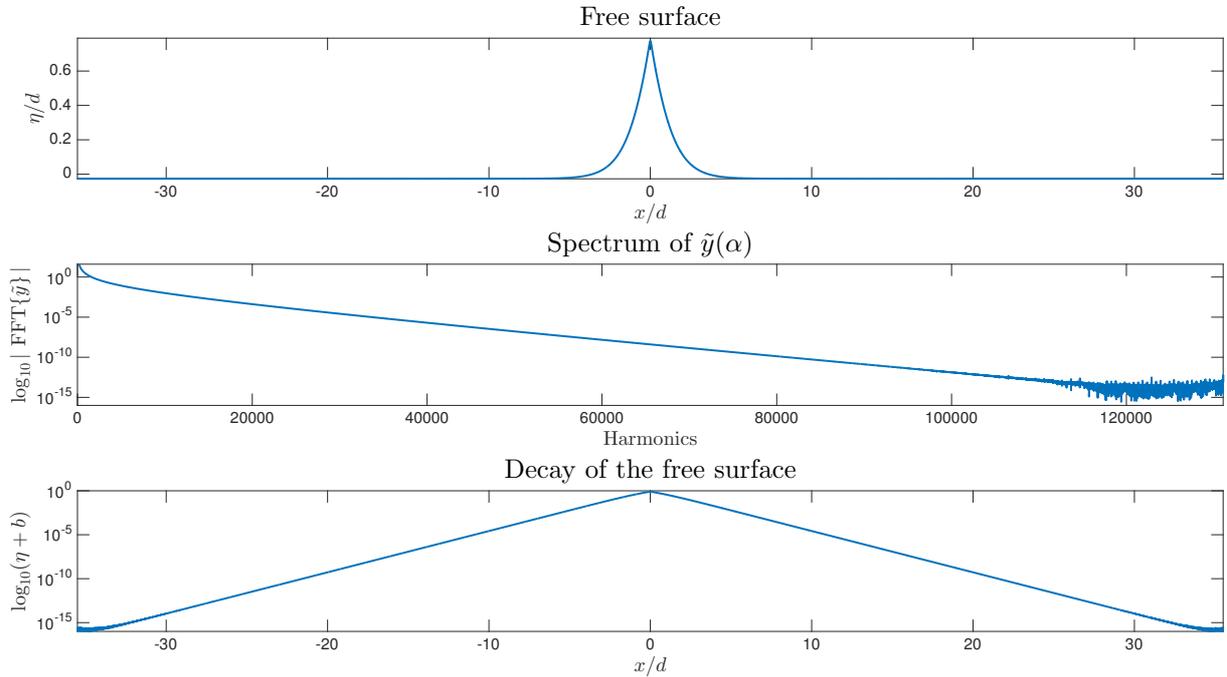}
  \caption{\small\em Steep cnoidal wave ($L/d\ =\ 71\,$, $H/d\ =\ 0.802$).}
  \label{figsolisurf}
\end{figure}


\subsection{Shallow water}
\label{sseccno}

When the depth $d$ over wavelength $L\ =\ 2\pi/k$ ratio is very small, \ie $kd\ll\ 1\,$, we are dealing with the so-called {\em shallow water\/} situation. It is well-known that \textsc{Stokes}' expansion fails to approximate such solutions of finite amplitude. Some shallow water approximations have then be proposed to approximate periodic waves, the so-called {\em cnoidal waves}. Many numerical algorithms devoted to the resolution of the full equations also fail in shallow water. For instance, \textsc{Fenton}'s algorithm \cite{Fenton1988} works only for $L/d\ \lesssim\ 30\,$.\footnote{For $L/d\ >\ 30\,$, \textsc{Fenton}'s algorithm does not converge or converge to ghost solutions with spurious oscillations, see Fig.~3-2 in \cite{Fenton2015}.}

The present algorithm works in shallow water without difficulties. For instance, with $L/d\ =\ 1000$ and $H/d\ =\ 0.4$ (\ie, $\varepsilon\ \approx\ 0.00126$ and $kd\ \approx\ 0.00628$) the solution is obtained (in about $0.4\,\mathsf{s}$ on a MacBook Pro laptop computer from 2012) using $N\ =\ 8\,192$ \textsc{Fourier} modes that are necessary to achieve machine double precision (Figure~\ref{figcnoisurf}). Another more extreme example is the case $L/d\ =\ 
10000$ and $H/d\ =\ 0.7\,$, that is computed in double precision with $N\ =\ 2^{\,19}$ and $\mathsf{tol}\ =\ 10^{\,-14}$ after $237$ iterations in about $1\,\mathsf{mn}$. It should be noted that these solutions are obtained from the initial guess \eqref{inigues} that is not at all a decent approximation of the solution, thus illustrating the robustness of the algorithm.

As the steepness or the wavelength increases, the number of \textsc{Fourier} modes $N$ required to achieve spectral accuracy increases rapidly. For example, with $L/d\ =\ 71$ and $H/d\ =\ 0.802$ (\ie, $\varepsilon\ \approx\ 0.0355$ and $kd\ \approx\ 0.0885$) a steep cnoidal wave is obtained (Figure~\ref{figsolisurf} upper), its computation to full spectral accuracy requiring $N\ =\ 2^{\,17}\ =\ 131\,072$ \textsc{Fourier} modes (Figure~\ref{figsolisurf} middle).\footnote{This result was obtained in less than three minutes. A rough estimate suggests that the same computation with \textsc{Newton} or \textsc{Levenberg--Marquardt} iterations, instead of \textsc{Petviashvili}'s ones, would take several days (possibly weeks) on the same computer.}

For steep cnoidal waves in very shallow water the number of necessary \textsc{Fourier} modes can be prohibitively large. An alternative is to compute a shorter cnoidal wave and to eventually increase the length of the trough. Indeed, a cnoidal wave surface decaying rapidly from the crest, it rapidly reaches its minimum to machine precision (Figure~\ref{figsolisurf} lower). The example of Figure~\ref{figsolisurf} shows that longer cnoidal waves can be obtained to machine precision increasing the length of the trough up to the desired wavelength, then redefining the mean water level and the mean depth, as well as all the related parameters (renormalisation). A similar procedure can be used to compute solitary waves, as shown below.

\begin{table}[htp]
\begin{center}
\begin{tabular}{|c|c|c|c|c|c|}
\hline
$L/d$\rule{0mm}{4mm} & $100$  & $1\,000$ & $10\,000$ & $100\,000$ & $1\,000\,000$ \\
\hline
$1-m$\rule{0mm}{4mm} & $2.04\times10^{-11}$ & $1.85\times10^{-118}$ & $6.91\times10^{-1189}$ 
& $3.63\times10^{-11893}$ & $5.78\times10^{-118936}$\\
\hline
\end{tabular}
\bigskip
\caption{\small\em KdV parameter $m$ for $H/d\ =\ 0.1\,$.}\label{tabHd01}
\end{center}
\end{table}

\begin{table}[htp]
\begin{center}
\begin{tabular}{|c|c|c|c|c|c|c|}
\hline
$L/d$\rule{0mm}{4mm} & $50$  & $60$ & $70$ & $80$ & $90$ & $100$ \\
\hline
$1-m$\rule{0mm}{4mm} & $8.06\times10^{-13}$ & $1.77\times10^{-15}$ & $3.87\times10^{-18}$ 
& $8.47\times10^{-21}$ & $1.86\times10^{-23}$ & $4.06\times10^{-26}$ \\
\hline
\end{tabular}
\bigskip
\caption{\small\em KdV parameter $m$ for $H/d\ =\ 0.5\,$.}\label{tabHd05}
\end{center}
\end{table}


\subsection{Comparison with KdV cnoidal wave}
\label{ssecKdV}

\textsc{Korteweg} and \textsc{de Vries} \cite{KdV} proposed an analytic approximation for small amplitude long periodic waves in shallow water. They coin the term `cnoidal' wave because this approximation can be expressed in term of the \textsc{Jacobi} cn$-$function. In our notations, KdV analytic solution can be conveniently written
\begin{gather}
  \eta\ =\ a\,\frac{K\/\operatorname{dn}^2\!\left(\/\kappa\/x\/|\/m\/\right)-\,E}{K\,-\,E}, \qquad
  k\ =\ \frac{\pi\,\kappa}{K}, \qquad H\ =\ \frac{m\,K\,a}{K\,-\,E}, \qquad  
  (\kappa\/d)^2\ =\ \frac{3\,H}{4\,m\,d}\,, \label{cnoKdV}
\end{gather}
$\text{dn}$ being the elliptic functions of \textsc{Jacobi} of parameter $m$ ($0\ \leqslant\ m\ \leqslant\ 1$), $K\ \eqdef\ \text{K}\/(m)$ and $E\ \eqdef\ \text{E}\/(m)$ being the complete elliptic integrals of the first and second kinds, respectively \cite{Abramowitz1965}.

Though an analytic approximation, KdV cnoidal wave requires significant computations. Indeed, a wave being generally defined for given height $H$ and wavenumber $k$, the parameter $m$ must be determined solving numerically the equations in \eqref{cnoKdV} relating the parameters. For very long waves, $m$ is very close to one, to an extend that it cannot be practically computed (see Tables~\ref{tabHd01} and \ref{tabHd05}). For instance, for the very long small amplitude cnoidal wave with $L/d\ =\ 10^{\,6}$ and $H/d\ =\ 10^{\,-1}$ we have $1\ -\ m\ \approx\ 5.78\times10^{\,-118936}\,$, a value that cannot be easily computed. This problem becomes more severe as the amplitude increases (Table~\ref{tabHd05}). Thus, for very long waves, KdV analytic cnoidal solution is useless for practical applications, even if only crude approximations are sufficient. It is then more efficient to solve numerically the KdV equation, for instance with FFT and \textsc{Petviashvili}'s method as illustrated here for the \textsc{Babenko} equation. But, doing so, solving KdV is not much less demanding than solving Babenko, so the latter should be preferred. For the extreme example\footnote{This example is given to illustrate the efficiency of the method, not for its practical interest.} $H/d\ =\ 10^{\,-1}$ and $L/d\ =\ 10^{\,6}\,$, \textsc{Babenko} solution is computed to double-precision spectral precision with $N\ =\ 2^{\,21}$ in about $100\,\mathsf{s}$ using our algorithm,\footnote{On the same hardware, for the same wave, rough estimates indicate that algorithms of complexity $\O\,(N^{\,2})$ would take months, and those of complexity $\O\,(N^{\,3})$ would take thousands of years.} while KdV analytic solution cannot be computed in double precision.\footnote{For a direct determination of $m$ from the relations \eqref{cnoKdV}, one would have to use something like two hundred thousand digits computation. An alternative is to derive better conditioned relations via some non trivial mathematical manipulations, thus loosing the analytical simplicity of KdV.} Of course, this drawback is not limited to the KdV analytic solution, it is {\em a fortiori} present in all cnoidal-like approximations, such as the solutions of the \textsc{Boussinesq}-like equations. These considerations demonstrates that our algorithm for solving the irrotational \textsc{Euler} equations is also a suitable alternative to simple analytic models.


\subsection{Solitary waves}
\label{ssecsol}

Solitary waves decaying exponentially fast, their surface elevation reaches zero to machine precision close to the crest. Thus, solitary waves can be efficiently computed in a periodic box, provided that the box is long enough so the periodisation does not affect the solution. This numerical trick is well-known and has been used by many authors. 

The steep cnoidal example of Figure~\ref{figsolisurf} reaches its minimal elevation to machine precision before its trough at $x\ =\ L/2$ (see Figure~\ref{figsolisurf} lower). Thus, this cnoidal wave can be considered as a solitary wave computed in a periodic box, but with a different still water level. The actual depth for the solitary wave is $d_\infty\ \eqdef\ d\ +\ \eta\,(L/2)\ =\ d\ -\ b\,$, the surface elevation is $\eta_\infty\ =\ \eta\ +\ b$ and the dimensionless amplitude is $H/d_\infty\,$. For the example of Figure~\ref{figsolisurf} we obtain $H/d_\infty\ =\ 0.8236847804878956\,$. This result is surprising because the direct computation of solitary waves via the CPM converges only for $H/d\ \lesssim\ 0.79$ \cite{Dutykh2013b, Clamond2012b}. The solitary wave thus obtained has been compared to an approximation obtained with \textsc{Tanaka}'s method using $1\,024$ nodes (\textsc{Tanaka}'s algorithm \cite{Tanaka1986} is way too slow to use it with $2^{\,18}$ nodes). We found that the two solutions match up to about five digits, that is consistent with the accuracy of \textsc{Tanaka}'s method \cite{Dutykh2013b, Clamond2012b}, confirming the solution obtained by the CPM after renormalisation.

\begin{figure}
  \centering
  \vspace*{0.80em}
  \includegraphics[width=0.99\textwidth]{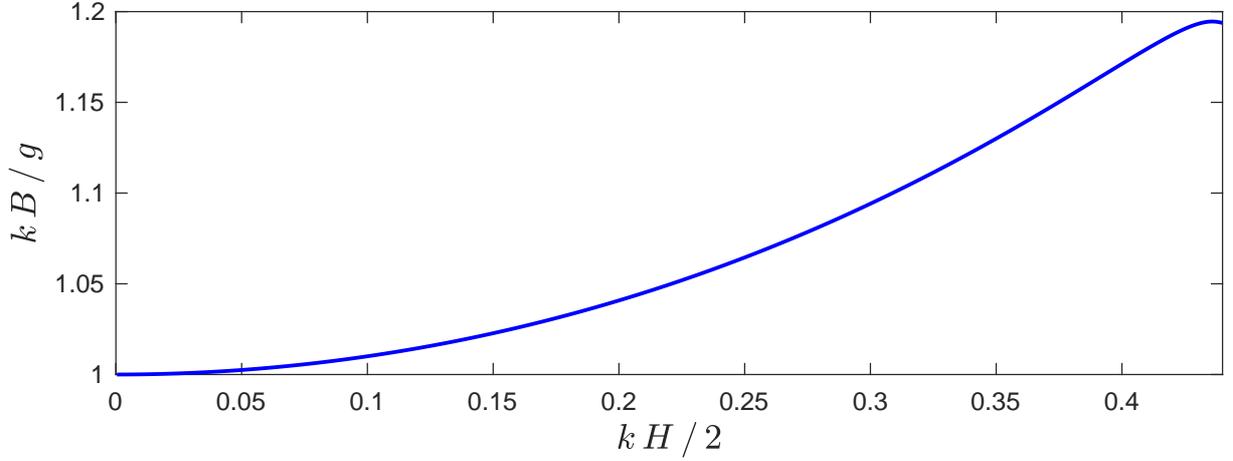}
  \caption{\small\em \textsc{Bernoulli} constant in deep water as function of the steepness.}
  \label{figBvsEPS}
\end{figure}

\begin{table}
\begin{center}
\begin{tabular}{ccccccc}
$kd$ & $\hat{\varepsilon}$  & $\varepsilon_{\mbox{\tiny\sc b}}$ & $\varepsilon_{\mbox{\tiny\sc c}}$ 
& $A(\varepsilon_{\mbox{\tiny\sc c}})$ & $\vartheta(\varepsilon_{\mbox{\tiny\sc c}})$ 
& $\theta_{\text{max}}(\varepsilon_{\mbox{\tiny\sc c}})$ \\ 
\hline
$\infty$ & $0.443164$ & $0.435907$ & $0.4400$ & $2.4366$ & $2.4710$ & $29.831^\circ$ \\
$2.0$ & $0.422293$ & $0.415166$ & $0.4199$ & $2.5651$ & $2.5865$ & $29.942^\circ$ \\
$1.5$ & $0.389554$ & $0.382626$ & $0.3876$ & $2.6336$ & $2.6475$ & $29.979^\circ$ \\
$1.0$ & $0.315872$ & $0.309415$ & $0.3146$ & $2.7535$ & $2.7573$ & $30.042^\circ$ \\
$0.9$ & $0.294147$ & $0.287847$ & $0.2930$ & $2.7714$ & $2.7734$ & $30.050^\circ$ \\
$0.8$ & $0.269865$ & $0.263760$ & $0.2688$ & $2.7656$ & $2.7674$ & $30.041^\circ$ \\
$0.7$ & $0.243084$ & $0.237227$ & $0.2421$ & $2.7525$ & $2.7547$ & $30.028^\circ$ \\
$0.6$ & $0.213965$ & $0.208426$ & $0.2130$ & $2.6964$ & $2.7007$ & $29.981^\circ$ \\
$0.5$ & $0.182750$ & $0.177626$ & $0.1818$ & $2.6216$ & $2.6297$ & $29.910^\circ$ \\
$0.4$ & $0.149709$ & $0.145132$ & $0.1488$ & $2.5396$ & $2.5520$ & $29.822^\circ$ \\
$0.3$ & $0.115019$ & $0.111166$ & $0.1142$ & $2.4542$ & $2.4723$ & $29.714^\circ$ \\
$0.2$ & $0.078662$ & $0.075763$ & $0.0780$ & $2.3636$ & $2.3886$ & $29.579^\circ$ \\
$0.106814$ & $0.043100$ & $0.041362$ & $0.0427$ & $2.3099$ & $2.3399$ & $29.488^\circ$ \\
\hline
\end{tabular}
\end{center}
\bigskip
\caption{\small\em Various parameters for the highest computable waves.}
\label{tabHigh}
\end{table}


\section{Remarks on the highest computable waves}
\label{sseclim}

In order to characterise the highest computable waves, we consider several dimensionless parameters. For a given $kd\,$, we denote $\hat{\varepsilon}\,$, $\varepsilon_{\mbox{\tiny\sc b}}$ and $\varepsilon_{\mbox{\tiny\sc c}}$ the steepnesses corresponding of, respectively, the highest wave (with a $120^\circ$ inner angle at the crest), the first maximum of the \textsc{Bernoulli} constant $B$ and the highest computable wave. We also consider the maximum inclination of the free surface $\theta_{\,\text{max}}$ and the parameters   
\begin{equation*}
  \vartheta\ \eqdef\ -\/\half\,\log\!\left(\,1\,-\,\varepsilon\,/\,\hat{\varepsilon}\,\right), \qquad 
  A\ \eqdef\ \log\!\left(\,\tilde{u}_1\,/\,\tilde{u}_0\,\right),
\end{equation*}
where $\tilde{u}_{\,0}$ and $\tilde{u}_{\,1}$ are the horizontal velocities at the crest and at the trough, respectively. Note that $\vartheta\ \to\ \infty$ and $A\ \to\ \infty$ as $\varepsilon\ \to\ \hat{\varepsilon}\,$, a wave is then considered steep if these parameters are larger than two \citep{Maklakov2002}.

The maximum computable steepness $\varepsilon_{\mbox{\tiny\sc c}}$ is determined (by dichotomy) for each $kd$ only up to the fourth decimal place, that is sufficient here for the discussion. All the computations of $\varepsilon_{\mbox{\tiny\sc c}}$ in Table~\ref{tabHigh} were performed in double precision, with $N\ =\ 131\,072$ \textsc{Fourier} positive modes and with tolerance $\mathsf{tol}\ =\ 10^{\,-12}$ for the iterations. The corresponding parameters $A\,$, $\vartheta$ and $\theta_{\text{max}}$ are given by truncated decimal expansions (\ie, not rounded to the nearest decimal approximation). In the Table~\ref{tabHigh}, the steepnesses of the highest waves $\hat{\varepsilon}$ and the steepnesses $\varepsilon_{\mbox{\tiny\sc b}}$ corresponding to the first maximum of the \textsc{Bernoulli} constants were kindly provided by Professor Dmitri \textsc{Maklakov} who guaranteed that the six decimals are correct. The values of $\varepsilon_{\mbox{\tiny\sc b}}$ were confirmed by our algorithm, thus providing another validation of the method.

As shown in Table~\ref{tabHigh}, the highest computable waves are rather steep with $\theta_{\text{max}}\ \approx\ 30^\circ\,$, $A$ and $\vartheta$ being significantly larger than two. More interestingly, $\varepsilon_{\mbox{\tiny\sc c}}$ always exceeds $\varepsilon_{\mbox{\tiny\sc b}}\,$, $\varepsilon_{\mbox{\tiny\sc c}}$ corresponding at least to $99\%$ of the maximum steepness $\hat{\varepsilon}\,$. This is somewhat surprising because similar algorithms for periodic waves in deep water \cite{Dyachenko2014} and solitary waves \cite{Dutykh2013b, Clamond2012b} have maximum computable steepnesses $\varepsilon_{\mbox{\tiny\sc c}}\ \approx\ \varepsilon_{\mbox{\tiny\sc b}}\,$.

Indeed, in their implementation of the GPM for the \textsc{Babenko} equation \eqref{babYter} with $\delta\ =\ 0\,$, \cite{Dyachenko2014} found convergence for $\varepsilon\lesssim0.436$, and for solitary waves the present authors \cite{Dutykh2013b, Clamond2012b} noticed that the CPM converges for $H/d\ \lesssim\ 0.79\,$. These maximum computable steepnesses correspond to the first maximum of the \textsc{Bernoulli} constant (Figure~\ref{figBvsEPS}, Table~\ref{tabHigh}). The limiting value  $\varepsilon_{\mbox{\tiny\sc c}}\ \approx\ \varepsilon_{\mbox{\tiny\sc b}}$ was then explained by the fact that, for steeper waves, the lack of one-to-one correspondance between the parameters prevents the algorithm to `decide' which solution should be retained.

The situation is a bit different here, where we found that systematically $\varepsilon_{\mbox{\tiny\sc c}}\ >\ \varepsilon_{\mbox{\tiny\sc b}}$ (see Table~\ref{tabHigh}), even for solitary waves (see Section~\ref{ssecsol}). Clearly, the modifications and rescaling, introduced to make the CPM work for arbitrary wavelengths, have a beneficial effect on the steepest computable waves. A possible explanation is that the rescaling somehow enlarge the region of one-to-one correspondance between the computed parameters. This is a conjecture requiring rigorous mathematical investigations that are far beyond the purpose of the present paper.


\section{Discussion}
\label{secconc}

We described an efficient algorithm for computing steady surface gravity waves for an ideal homogeneous fluid in irrotational motion. After analytic transformations (conformal mapping, rewriting of the conditions at the free surface, change of dependent variables) we ended up with a modified \textsc{Babenko}-like equation that can be solved numerically via the classical \textsc{Petviashvili} method (CPM). The algorithm thus obtained is very fast --- with complexity $\O\,(N\log N)\,$, $N$ being the number of \textsc{Fourier} modes --- and any accuracy can be reached for all depths, provided that $N$ is large enough and that the steepness is not too close to the limiting one. All waves of practical interest can therefore be computed. To our knowledge, it is the first algorithm uniformly valid for all wavelength-over-depth ratios, in practice and not only in principle, that is moreover accurate in the sense that arbitrary precision numerical solutions can be obtained.

With this algorithm, the computation of steady gravity waves for the irrotational \textsc{Euler} equation is not more demanding than the numerical resolution of simplified models such as KdV. It should be noted that some simplified water wave models --- \eg, some variants of the \textsc{Boussinesq} equations for long waves in shallow water --- have inhomogeneous nonlinear term. Therefore, the (like of) \textsc{Petviashvili} method does not work and \textsc{Newton} or \textsc{Levenberg--Marquardt} methods should be used instead. The latter having complexity $\O\,(N^{\,3})\,$, the numerical resolution of these simplified models is more demanding than our algorithm for the irrotational \textsc{Euler} equations.

It is often believed that the CPM works only for localised (solitary) waves. Here, we disproved this belief with strong numerical evidences. Rigorous mathematical results are scarce \cite{Alvarez2014, Pelinovsky2004} for the (like of) \textsc{Petviashvili} method. Deeper mathematical understanding would be beneficial for improving the method, in particular for the computation of almost highest waves. We hope that the numerical evidences presented here will stimulate such investigations.

There exist steady surface gravity waves with different crests and asymmetric waves \cite{Okamoto2001}. However, these solutions exist close to the limiting ones where the classical \textsc{Petviashvili} method does not work, at least not as formulated in this paper. Improving this fixed point iteration method may increase the range of computable steepnesses. However, this will not be sufficient for the computation of extreme waves using \textsc{Fourier} decomposition together with the conformal mapping, because a huge number of \textsc{Fourier} modes is needed to achieve high accuracy. This can be understood considering the limiting wave with an $120^\circ$ inner angular crest. Such solutions have a power $2/3$ singularity at the crest in the conformal plane \cite{Stokes1880a} and, therefore, the $n$-th Fourier coefficient decays like $n^{-5/3}$ as $n\ \to\ \infty\,$. Thus, truncating the \textsc{Fourier} expansion after the $N$\up{th} term, the error decays like $N^{\,-2/3}$ as $N\ \to\ \infty\,$. In practice, it means that in order to increase the accuracy by two digits, the number of computed \textsc{Fourier} modes must be multiplied by (roughly) one thousand! Clearly, the `brutal force' approach consisting in massively increasing $N$ is inefficient for extreme waves, even on a powerful computer. To overcome this difficulty, several authors have proposed a change of independent variable such that the corresponding \textsc{Fourier} spectrum decays faster. Another possibility for accurate computations of extreme waves consists in a totally different mathematical reformulation of the problem, so that the numerical resolution is less demanding, as suggested in \cite{Clamond2018a}. Needless to say, such approaches are suitable if the number of modes $N$ required for high accuracy is relatively small if the algorithmic complexity is $\O\,(N^{\,2})$ or $\O\,(N^{\,3})\,$, or if the algorithmic complexity remains $\O\,(N\log N)$ with $N$ not too large. To our knowledge, an efficient algorithm for arbitrary precision calculation of extreme waves has yet to be discovered.

Here, we introduced simple tricks in order to successfully apply the classical \textsc{Petviashvili} method for computing steady water waves of arbitrary wavelength, in arbitrary depth and to arbitrary precision. These tricks can certainly be used for other equations involved in fluid mechanics.


\subsection*{Acknowledgments}
\addcontentsline{toc}{subsection}{Acknowledgments}

This work was supported by the \textsc{Spanish} {\em Ministerio de Econom\'{\i}a y Competitividad\/} under the Research Grant MTM2014-54710-P. The authors would like to thank Professor Angel \textsc{Dur\'an} for useful discussions. The authors are also grateful to Professor Dmitri~\textsc{Maklakov} for providing some valuable data in the Table~\ref{tabHigh} and whose comments helped to improve the paper.


\appendix
\section{Potential and stream function in a `fix' frame of reference}
\label{apppsi}

The definition \eqref{defce} of $\ce$ implies that the function $\Phi\ \eqdef\ \phi\ +\ \ce\/x$ averages zero along any horizontal line $y\ =\ \text{constant}\,$, in particular at the bed $y\ =\ -\depth\,$. This interesting property suggests the introduction of a stream function $\Psi$ in this frame of reference where the mean horizontal velocity is zero at the bed and such that
\begin{equation*}
  \Psi\ \eqdef\ \psi\ -\ \bot{\psi}\ +\ \ce\,(y+\depth), \qquad
  \bot{\Psi}\ =\ 0, \qquad \sur{\Psi}\ =\ \ce\,(\eta+\depth)\ -\ \cs\,\depth,
\end{equation*}
thence
\begin{equation*}
  \left<\,\Psi(x,y=\text{constant})\,\right>\,=\ 0, \qquad 
  \left<\,\sur{\Psi}(x)\,\right>\,=\,\left<\,\Psi(x,y=\eta(x))\,\right>\,=\ (\ce-\cs)\,\depth\,.
\end{equation*}
The definitions of $\ce$ and $\cs$ also yield (integrating by parts and exploiting the irrotationality)
\begin{align*}
  \left<\,\sur{\Psi}\,\right>\,&=\,\left<\,\int_{-\depth}^\eta u\,\ud\/y\ +\ \ce\,\depth\,\right>\,=\,\left<\,\int_{-\depth}^\eta u\,\ud\/y\ -\ \depth\,\bot{u}\,\right>\,=\,\left<\,\eta\,\sur{u}\ -\, \int_{-\depth}^\eta y\,u_y\,\ud\/y\,\right>\nonumber\\
  &=\,\left<\,\eta\,\sur{u}\ -\, \int_{-\depth}^\eta y\,v_x\,\ud\/y\,\right>\,=\,\left<\,\eta\,\sur{u}\ +\ \eta\,\eta_x\,\sur{v}\ -\ \frac{\partial}{\partial\/x}\int_{-\depth}^\eta y\,v\,\ud\/y\,\right>\,=\,\left<\,\eta\,\sur{\phi}_x\,\right>\,,
\end{align*}
thence 
\begin{equation*}
  \left<\,\eta\,\sur{\phi}_x\,\right>\,=\ (\ce-\cs)\,\depth\,, 
\end{equation*}
that is exploited in the relation \eqref{meanlevelCM}. It should be noticed that the quantity $(\cs\ -\ \ce)\/\depth$ (related to the wave impulse, see Appendix~\ref{appint}) must remain bounded as $\depth\to\infty$, implying that $\cs\to\ce$ as $\depth\to\infty\,$. Note also that, the velocity magnitude varying monotonically  from the bottom to the surface, we necessarily have $\ce\ \geqslant\ \cs$ \cite{Constantin2013} and therefore $\left\llangle\sur{y}\right\rrangle\ \leqslant\ 0$ (see Eq.~\ref{meanlevelCM}).


\section{Integral quantities}\label{appint}

The wave can be characterised by several integral parameters \cite{Longuet-Higgins1975, Longuet-Higgins1984, McCowan1891, Starr1947}. In the frame of reference moving with the wave, there are three physically important constants: the fluid flow $\mathcal{Q}_{\,0}\,$, the momentum flux $\mathcal{S}_{\,0}$ and the energy flux $\mathcal{F}_{\,0}\,$, defined by
\begin{align}
\mathcal{Q}_{\,0}\ &\eqdef\,\int_{-\depth}^\eta\,u\,\ud\/y\ =\ \sur{\psi}\ -\ 
\bot{\psi}\ =\ -\,\cs\,\depth, \label{defmasflux} \\
\mathcal{S}_{\,0}\ &\eqdef\, \int_{-\depth}^\eta\left(\/p\/+\/u^{\,2}\/\right)\ud\/y\ =\,\left[\left(\/p\/
+\/u^{\,2}\/\right)(y+\depth)\,\right]_{-\depth}^{\eta}\ -\ \int_{-\depth}^\eta\left(\/p_{\,y}\/
+\/2\/u\/u_y\/\right)(y+\depth)\,\ud\/y\nonumber\\
&=\ \sur{u}^2\,(\eta+\depth)\ +\ \int_{-\depth}^\eta\left(\/g\/-\/u\/v_x\/-\/v\/u_x\/\right)(y+\depth)
\,\ud\/y\nonumber\\
&=\,\left(\sur{u}^{\,2}+\sur{v}^{\,2}\right)(\eta+\depth)\ +\ \half\,g\,(\eta+\depth)^{\,2}\ -\ \partial_x\!\int_{-\depth}^\eta\,
u\,v\,(y+\depth)\,\ud\/y \nonumber\\
&=\ B\,(\eta+\depth)\ +\ 2\,g\,\depth\,(\eta+\depth)\ -\ \threehalf\,g\,(\eta+\depth)^{\,2}\ -\ \partial_x\!\int_{-\depth}^\eta\,u\,v\,(y+\depth)\,\ud\/y,\\
\mathcal{F}_{\,0}\ &\eqdef\,\int_{-\depth}^\eta\left[\,p\,+\,\half\,u^{\,2}\,+\,\half\,v^{\,2}\,+\,g\,y\,\right]u\,\ud\/y\ =\ \int_{-\depth}^\eta\half\,B\,u\,\ud\/y\ =\ \half\,B\,\mathcal{Q}_{\,0}\,. \label{defeneflux}
\end{align}
These quantities are related to other averaged quantities of physical interest (see below). In particular, averaging $\mathcal{S}_{\,0}$ over one wavelength and exploiting the relation \eqref{defBot} and the impermeability of the free surface, one obtains at once
\begin{equation*}
  \mathcal{S}_{\,0}\ =\ B\,\depth\ +\ \half\,g\,\depth^{\,2}\ -\ 3\,\mathcal{V}\,, 
\end{equation*}
where $\mathcal{V}\ \eqdef\ \left<\,\half\,g\,\eta^{\,2}\,\right>$ is the potential energy of the gravity force.

Other integral quantities can be defined relatively to the uniform flow of speed $-\cl\,$, \ie, in the fixed frame of reference where the phase velocity is $\cl\,$. The integral quantities of interest here are the:
\begin{align}
\renewcommand{\arraystretch}{1.7}
\begin{array}{rrlc}
\text{\small Circulation:}& \mathcal{C} \!&\! \eqdef\,\left<\,\sur{u}\,+\,\cl\,+\,\sur{v}\,\eta_x\,\right>
\,=\ \cl\ -\ \ce, \\
\text{\small Impulse:}& \mathcal{I} \!&\! \eqdef\,\left<\,\int_{-\depth}^\eta\,
(u+\cl)\,\ud\/y\,\right>\, =\,\left(\,\cl\,-\,\cs\,\right)\depth, \\
\text{\small Kinetic Energy:}& \mathcal{K} \!&\! \eqdef\,\left<\,\int_{-\depth}^{\eta}
\half\,[\,(u+\cl)^{\,2}+v^{\,2}\,]\,\ud\/y\,\right>\, =\ \half\,\cl\,\mathcal{I}\ -\ \half\,\depth\,
\cs\,\mathcal{C},  \\
\text{\small Radiation Stress:}& \mathcal{S}_{xx} \!&\! \eqdef\,\left<\,\int_{-\depth}^\eta
\left[\,p\,+\,(u+\cl)^{\,2}\,+\,g\,y\,\right]\ud\/y\,\right>\,=\ 2\,\cl\,\mathcal{I}\ -\ 
2\,\mathcal{V}\ +\, \left(B-\cl^{\,2}\right)\depth, \\
\text{\small Momentum Flux:}& \mathcal{S} \!&\! \eqdef\,\left<\,\int_{-\depth}^\eta\left[\,
p\,+\,(u+\cl)^{\,2}\,\right]\ud\/y\,\right>\,
=\ \mathcal{S}_{xx}\ -\ \mathcal{V}\ +\ \half\,g\,\depth^2, \\
\text{\small Energy Flux:}& \mathcal{F} \!&\! \eqdef\,\left<\,\int_{-\depth}^\eta\left[\,
p\,+\,\half\/(u+\cl)^{\,2}\,+\,\half\/v^{\,2}\,+\,g\/y\,\right](u+\cl)\,\ud\/y\,\right>\nonumber\\
& &\!=\ \half\left(\/B\/-\/\cl^{\,2}\/\right)\cl\depth\ +\ \half\left(\/B\/+\/
\cl^{\,2}\/\right)\mathcal{I}\ +\,\left(\mathcal{K}-2\mathcal{V}\right)\cl,\\
\text{\small Group celerity:}& c_g \!&\! \eqdef\ \mathcal{F}\,/\,(\mathcal{K}+\mathcal{V}).
\end{array}
\renewcommand{\arraystretch}{1.0}
\end{align}
The equalities in these integral relations are easily obtained via some trivial derivations. Note that the radiation stress defined here differs from the definition of \cite{Longuet-Higgins1975}, that is $\mathcal{S}_{\,x\,x}^{\,\text{\tiny\sc LH}}\eqdef\mathcal{S}_{\,x\,x}\ -\ \mathcal{V}\,$. Note also that the group celerity defined above is not the linear one, \ie $c_g\ \neq\ \partial(k\,c_{\,0})/\partial k$ if $H\ \neq\ 0\,$.


\section{Relations between averaged quantities in physical and conformal planes}\label{appaverel}

Averaging in the physical plane is different than averaging in the conformal plane. For practical applications, many averaged quantities may have to be computed. For an easy reference, we give below various connections between averaged quantities in the physical and conformal planes. In particular, we have the special ``conformal averaged'' relations at the free surface
\begin{align}
\left\llangle\sur{y}\right\rrangle\,&=\ -\,\ce^{\,-1}\left<\,\eta\,\sur{\phi}_x\,
\right>\,=\,\left(\cs\,\ce^{-1}\,-\,1\right)\depth, \label{meanlevelCM} \\
\left\llangle\,\mathscr{C}\!\left\{\sur{y}\right\}\,\right\rrangle\,&=\ \cl\,\ce^{\,-1}\ 
-\ \cl\,\cs^{\,-1}\ =\ \cl\,\cs^{\,-1}\,\depth^{-1}\left\llangle\,\sur{y}\,\right\rrangle, \\
\left\llangle\,\mathscr{C}^{-1}\!\left\{\sur{y}\right\}\,\right\rrangle\,&=\,\left(\,
\cs\,\ce^{\,-1}\,-\,1\,\right)\cs\,\cl^{-1}\,\depth^{\/2}\ =\ \cs\,\cl^{-1}\,\depth 
\left\llangle\,\sur{y}\,\right\rrangle, \label{meanCieCM}\\
\left\llangle\,\sur{y}\,\mathscr{C}\!\left\{\sur{y}\right\}\,\right\rrangle\,
&=\,\left(\,\ce\,-\,\cs\,\right)\cl\,\ce^{\,-1}\,\cs^{\,-1}\,\depth\ =\ -\,\cl\,\cs^{-1}
\left\llangle\,\sur{y}\,\right\rrangle,\\
\left\llangle\,\mathscr{C}^{-1}\!\left\{\sur{y}\,\mathscr{C}\!\left\{\sur{y}\right\}\right\}\,
\right\rrangle\,&=\,\left(\,1\,-\,\cs\,\ce^{\,-1}\,\right)\depth^{\/2}\ =\ -\,\depth
\left\llangle\,\sur{y}\,\right\rrangle, \label{meanCieCeCM} \\
\left\llangle\sur{u}\right\rrangle\,&=\ -\,\ce^{\,-1}\left<\,\sur{u}\,(\sur{u}+
\sur{v}\/\eta_x)\,\right>\,=\ -\,\ce^{\,-1}\left<\,\sur{u}^2+\sur{v}^2\,\right>\,
=\ -\,\ce^{\,-1}\,B,\\
\left\llangle\sur{v}\right\rrangle\,&=\ -\,\ce^{\,-1}\left<\,\sur{v}\,(\sur{u}+\sur{v}\/\eta_x)
\,\right>\,=\ -\,\ce^{\,-1}\left<\,\eta_x\left(\sur{u}^2+\sur{v}^2\right)\right>\,=\ 0,\\
\left\llangle\sur{u}^2+\sur{v}^2\right\rrangle\,&=\ B\ -\ 2\,g\left\llangle\sur{y}
\right\rrangle\,=\ B\ -\ 2\,g\left(\cs\,\ce^{-1}\,-\,1\right)\depth,\\
\left\llangle\sur{u}^{-1}\right\rrangle\,&=\ -\,\ce^{\,-1}\left<\,\sur{u}^{-1}\,
(\sur{u}+\sur{v}\/\eta_x)\,\right>\,=\ -\,\ce^{\,-1}\left<\,1\,+\,\eta_x^{\,2}\,\right>,\\
\left\llangle\sur{w}^{-1}\right\rrangle\,&=\ -\,\frac{1}{\ce}\left<\,\frac{\sur{u}\,+\,\sur{v}\,\eta_x}
{\sur{u}\,-\,\ui\,\sur{v}}\,\right>\,=\ -\,\frac{1}{\ce}\left<\,\frac{1\,+\,\eta_x^{\,2}}
{1\,-\,\ui\,\eta_x}\,\right>\,=\ -\,\frac{\left<\,1\,+\,\ui\,\eta_x\,\right>}{\ce}\,=\ -\,\frac{1}{\ce}, 
\end{align}
with $\sur{y}\,(\alpha)\ =\ \eta\,(\sur{x}\,(\alpha))$ and at the bottom
\begin{align}
\left\llangle\bot{y}\right\rrangle\,&=\ \ce^{\,-1}\left<\,\depth\,\bot{\phi}_x\,
\right>\,=\ -\/\depth, \\
\left\llangle\bot{u}\right\rrangle\,&=\ -\,\ce^{\,-1}\left<\,\bot{u}^2\,\right>\,
=\ -\,\ce^{\,-1}\,B,\\
\left\llangle\bot{u}^2\right\rrangle\,&=\ B\ -\ 2\,g\left\llangle\bot{y}\right\rrangle\,
-\ 2\,\left\llangle\bot{p}\right\rrangle\,=\ B\ +\ 2\,g\,\depth\ +\ 2\,\ce\left<\,
\bot{p}\,\bot{u}\,\right>,\\
\left\llangle\bot{u}^{-1}\right\rrangle\,&=\ -\/\ce^{\,-1},
\end{align}
with $\bot{y}=-\depth$. We also have the special ``physical averaged'' relations
\begin{align}
\left<x\right>\,&=\,\left\llangle\widetilde{x\,x_\alpha}\right\rrangle\,=\,\left\llangle
\Bot{x\,x_\alpha}\right\rrangle\,=\ \pi\,/\,k, \\
\left<\eta\right>\,&=\,\left\llangle\widetilde{y\,x_\alpha}\right\rrangle\,=\,
-\,\ce\left\llangle\,\sur{y}\,\sur{u}\,/\,(\sur{u}^2+\sur{v}^2)\,\right\rrangle\,=\ 0,\\
\left<\sur{u}\right>\,&=\ -\,\ce\left\llangle\,\sur{u}^2\,/\left(\sur{u}^2+\sur{v}^2\right)\,\right\rrangle\,
=\,\left<\,\sur{\phi}_x\,/\left(1+\eta_x^{\,2}\right)\,\right>,\\
\left<\sur{v}\right>\,&=\ -\,\ce\left\llangle\,\sur{u}\,\sur{v}\,/\left(\sur{u}^2+\sur{v}^2\right)\,\right\rrangle\,
=\,\left<\,\sur{\phi}_x\,\eta_x\,/\left(1+\eta_x^{\,2}\right)\,\right>,\\
\left<\sur{u}^2+\sur{v}^2\right>\,&=\,\left<\,\sur{\phi}_x^{\,2}\,/\left(1+\eta_x^{\,2}\right)\,\right>.
\end{align}
All these relations can be easily obtained from their definition and using the transformations \eqref{defmeanalphasur}--\eqref{defmeanxbot} between the averaging operators. Other relations can be similarly obtained.


\section{Velocity and pressure fields in the fluid}\label{appvel}

In the numerical procedure described below, we use conformal mapping and a \textsc{Fourier} pseudo-spectral method to solve the equations. This means that we obtain a discrete approximation equally spaced along each streamline. However, for practical applications, it is often necessary to determine the fields (velocity, pressure, \etc) at various positions that are not necessarily the nodes used for the computation. These informations can be obtained as follows.

Let be $W\,(z)\ \eqdef\ \ce\ +\ w\,(z)$ the complex velocity observed in the frame of reference where the fluid velocity averages to zero at the bottom. In this frame of reference, it follows that the complex potential $F\,(z) \eqdef \ce\,z\ +\ f(z)$ (\ie, $W\ =\ \ud F/\ud z$) is a periodic function, bounded in the whole fluid domain ($F$ is unbounded in any other frame of reference). 

The complex velocity being known at the free surface from our approximation procedure, $W$ at any complex abscissa $z$ can be obtained from the \textsc{Cauchy} integral
\begin{equation}\label{cauchint}
  \ui\,\theta\,W\,(z)\ =\ \mathrm{P.V.}\ointctrclockwise\frac{\ce\,+\,w(z_1)}{z_1\,-\,z}\,\ud\/z_{\,1}\,,
\end{equation}
where $\theta\ =\ 2\pi$ if $z$ is strictly inside the fluid domain (\ie, $\Im(z)\ <\ \eta$), $\theta\ =\ \pi$ if $z$ is at the free surface (\ie, $\Im(z)\ =\ \eta$) and $\theta\ =\ 0$ if $z$ is strictly above the free surface (\ie, $\Im(z)\ >\ \eta$). The bottom impermeability being taken into account via the method of images, the \textsc{Cauchy} integral \eqref{cauchint} yields for any $z$ below the free surface
\begin{align}\label{Wintinf}
  W\,(z)\ =\ \frac{\ui}{2\,\pi}\int_{-\infty}^{\infty}\left[\,\frac{\ce\,\sur{z}'(\alpha)\,-\,\cl}{\sur{z}(\alpha)\,-\,z}\ -\ \frac{\ce\,\sur{z}'^\ast(\alpha)\,-\,\cl}{\sur{z}^\ast(\alpha)\,-\,2\/\ui\/\depth\,-\,z}\,\right]\ud\/\alpha\,,
\end{align}
where $\sur{z}\,(\alpha)\ =\ (\cl/\ce)\alpha\ +\ \sur{X}(\alpha)\ +\ \ui\eta(\alpha)$ and $\sur{z}^{\,\prime}(\alpha)\ =\ \ud\sur{z}/\ud\alpha\ =\ (\cl/\cs)\ +\ \mathscr{C}\{\eta\}\,(\alpha)\ +\ \ui\eta_\alpha(\alpha)\,$, $\sur{X}$ and $\eta$ being known from the numerical resolution of the \textsc{Babenko} equation. 

The integral relation \eqref{Wintinf} is not suitable for periodic domains. For the latter, the infinite integral is replaced by one over one period involving the \textsc{Hilbert} kernel
\begin{align}\label{Wintper}
  W\,(z)\ =\ \frac{\ui\,k}{4\,\pi}\int_{-\pi/\kbar}^{\pi/\kbar}&\left\{\left[\,\ce\,\sur{z}'(\alpha)\,-\,\cl\,\right]\cot\!\left[\,\frac{k\,\sur{z}(\alpha)\,-\,k\,z}{2}\,\right]\right.\nonumber\\ 
  &\left.-\,\left[\,\ce\,\sur{z}'^\ast(\alpha)\,-\,\cl\,\right]\cot\!\left[\,\frac{k\,\sur{z}^\ast(\alpha)\,-\,2\,\ui\,k\,\depth\,-\,k\,z}{2}\,\right]\right\}\ud\/\alpha\,.
\end{align}
From this relation, we obtain the derivative of $W$ (required to compute the acceleration field)
\begin{align}
  \frac{\ud\,W(z)}{\ud\/z}\ =\ \frac{\ui\,k^2}{4\,\pi}\int_{-\pi/\kbar}^{\pi/\kbar}&
  \left\{\frac{\ce\,\sur{z}'(\alpha)\,-\,\cl}{1\,-\,\cos\!\left[\,k\,\sur{z}(\alpha)\,-\,k\,z\,\right]}\,-\,\frac{\ce\,\sur{z}'^\ast(\alpha)\,-\,\cl}{1\,-\,\cos\!\left[\,k\,\sur{z}^\ast(\alpha)\,-\,4\,\ui\,k\,\depth\,-\,k\,z\,\right]}\right\}\ud\/\alpha\,,
\end{align}
and the complex potential
\begin{align}
  F\,(z)\ =\ \int_{-\pi/\kbar}^{\pi/\kbar}&\left\{\,\frac{\ce\,\sur{z}'(\alpha)\,-\,\cl}{2\,\pi\,/\,\ui}\log\!\left[\frac{\sin\!\left(\half k(\sur{z}(\alpha)+\ui\depth)\right)}{\sin\!\left(\half k(\sur{z}(\alpha)-z)\right)}\right]\right.\nonumber\\
  &\left. +\ \left(\,\frac{\ce\,\sur{z}'(\alpha)\,-\,\cl}{2\,\pi\,/\,\ui}\log\!\left[\frac{\sin\!\left(\half k(\sur{z}(\alpha)+\ui\depth)\right)}{\sin\!\left(\half k(\sur{z}(\alpha)+2\ui\depth-z^*)\right)}\right]\,\right)^{\!\ast}\,\right\}\ud\/\alpha\,,
\end{align}
such that $W\ =\ \ud F/\ud z$ and $\Im(F)\ =\ 0$ at the bed.


\addcontentsline{toc}{section}{References}
\bibliographystyle{abbrv}
\bibliography{biblio}

\end{document}